\documentclass[authoryear]{elsarticle}

\journal{Advances in Water Resources}

\renewcommand{\vec}[1]{{\bf #1}}
\newcommand{\co}{CO$_{2}$ }
\newcommand{\coo}{CO$_{2}$}
\newcommand{\error}{$\mathrm{L}_{1}$ }
 \usepackage{color,graphicx,afterpage}

\begin{document}
\begin{frontmatter}

\title{Implicit Finite Volume and Discontinuous Galerkin Methods for Multicomponent Flow in Unstructured 3D Fractured Porous Media}

\author{Joachim Moortgat, Mohammad Amin Amooie, Mohamad Reza Soltanian}

\address{Corresponding author: J. Moortgat. School of Earth Sciences, the Ohio State University,
Columbus, Ohio, USA.}

\begin{abstract}
We present a new implicit higher-order finite element (FE) approach to efficiently model compressible multicomponent fluid flow on unstructured grids and in fractured porous subsurface formations. The scheme is sequential implicit: pressures and fluxes are updated with an implicit Mixed Hybrid Finite Element (MHFE) method, and the transport of each species is approximated with an implicit second-order Discontinuous Galerkin (DG) FE method. Discrete fractures are incorporated with a cross-flow equilibrium approach. 
This is the first investigation of all-implicit higher-order MHFE-DG for unstructured triangular, quadrilateral (2D), and hexahedral (3D) grids and discrete fractures. A lowest-order implicit finite volume (FV) transport update is also developed for the same grid types. The implicit methods are compared to an  Implicit-Pressure-Explicit-Composition (IMPEC) scheme. For fractured domains, the unconditionally stable implicit transport update is shown to increase computational efficiency by orders of magnitude as compared to IMPEC, which has a time-step constraint proportional to the pore volume of discrete fracture grid cells. However, when lowest-order Euler time-discretizations are used, numerical errors increase linearly with the larger implicit time-steps, resulting in high numerical dispersion. Second-order Crank-Nicolson implicit MHFE-DG and MHFE-FV are therefore presented as well. Convergence analyses show twice the convergence rate for the DG methods as compared to FV, resulting in two to three orders of magnitude higher computational efficiency. 
Numerical experiments demonstrate the efficiency and robustness in modeling compressible multicomponent flow on irregular and fractured 2D and 3D grids, even in the presence of fingering instabilities. 
\end{abstract}

\begin{keyword}
discontinuous Galerkin\sep discrete fractures\sep implicit higher-order methods \sep unstructured 3D grids \sep gravitational fingering \sep compressible multicomponent flow
\end{keyword}

\end{frontmatter}

\section{Introduction}
Fluid flow in fractured porous media is a challenging problem in reservoir simulations due to the wide range in spatial scales, rock properties, and resulting fluid velocities. In numerical simulations, if both fractures and the matrix are discretized by volumetric grid cells in the same mesh, all fracture-matrix interactions can theoretically be modeled with the same accuracy as for unfractured reservoirs (see, e.g., \citet{eikemo2009discontinuous} and references therein for an elegant approach based on the DG method). 
However, the single-continuum approach requires small time steps when explicit methods are used, while the large contrasts in permeabilities between neighboring grid cells can result in ill-conditioned matrices in implicit methods. 

The most popular alternative approach (and industry standard) has become the family of dual-continuum models, in which two homogenized grids are used: one for the fractures, which account for the flow, and one for the matrix, which accounts for the additional storage capacity (see \citet{warrenroot63, kazemi69, thomas83, arbogast90} for some of the early developments). This approach is computationally highly efficient, but relies on a simplified representation of the fracture-matrix interactions, which may be inadequate for certain complex flow problems.  

A third broad class of methods that is emerging is that of discrete fracture models, which strive to retain the accuracy of a single-continuum description, but to achieve higher computational efficiency \citep{noorishad1982upstream, baca1984modelling, granet1998single}. The fracture network is not homogenized and both the matrix and a network of discrete fractures, ideally at arbitrary orientations, are represented by grid cells that can theoretically accommodate all advective, diffusive, capillary, and gravitational fluxes. {\textit{Embedded} discrete fracture models have been proposed as well, which use common finite difference methods on structured grids, but account for fracture volumes by computing the intersections of fractures with the matrix grid (see, e.g., \cite{li2008efficient, moinfar2014development} and references therein).}

One class of discrete fracture models represents the fractures by ($D\!-\!1$) dimensional grid cells, where $D$ is the dimension. In terms of implementation, this is generally done by treating the edges (2D) or faces (3D) of $D$-dimensional matrix grid cells as a separate grid of fracture elements, for which pressures, velocities, and mass conservation can be solved separately or simultaneously with the matrix grid. This approach has been used in combination with control volume finite volume (FV) and finite element (FE) methods (e.g., \citet{bastian2000numerical, geiger2003combining, geiger2009black,karimifractures2004, monteagudo2004control, Monteagudo2007} and others) on 2D and 3D unstructured grids.  

Similar lower-dimensional fractures have also been modeled by a combination of Mixed-Hybrid FE (MHFE) methods for the pressure and flux fields, and higher-order discontinuous Galerkin (DG) methods, applied to single-phase \citep{martin2005modeling}, two-phase immiscible and incompressible flow \citep{jaffre2011discrete} and with capillarity on unstructured 3D grids \citep{hoteit2008}, and to single-phase compressible flow on 2D structured grids \citep{zidane2014efficient}. 

A different discrete fracture approach was developed for compositional and compressible multiphase flow \citep{hoteit2005}. In this so-called Cross-Flow-Equilibrium (CFE) method, fractures are embedded in $D$-dimensional grid cells, but it is assumed that the fractures instantaneously equilibrate with a small region of the neighboring matrix blocks. The fractures are combined with this neighborhood into larger computational grid cells (CFE elements), which avoids the explicit calculation of the fracture-matrix flux and results in a computationally efficient scheme. The flux contributions from the fracture and matrix portions of the CFE elements are integrated exactly by the MHFE method, while the mass conservation or transport update in the CFE elements is identical to that of the matrix grid cells. The transport update is approximated by a higher-order DG method in both the fractures and the matrix.
This approach was initially developed for two-phase flow on 2D structured grids \citep{hoteit2005} and has since been generalized to three-phase compositional flow on unstructured 2D and 3D grids and accounting for capillary pressure, gravity, and Fickian diffusion \citep{moortgatVII, moortgatVI, moortgatV}. Its predictions were recently compared to experiments and to a commercial dual-porosity simulator in \citet{moortgatVIII}.

A limitation of the aforementioned implementations has been that an implicit-pressure-explicit-composition (IMPEC) scheme was used for the MHFE- DG discrete CFE fracture model. The conditional stability of the explicit transport update incurs a CFL \citep{cfl1} constraint on the time-step size. The CFE elements can be much wider than the fracture aperture and in 2D IMPEC-CFE simulations are quite efficient. However, in 3D the pore-volume of grid cells, particularly at the intersection of multiple fractures, becomes significantly smaller than the matrix grid cells and the CFL condition can make the scheme prohibitively expensive. \citet{zidane2014efficient} mistook this as an inherent weakness of the CFE approach itself in comparing the computational efficiency of a different IMPEC-CFE implementation to an \textit{implicit} FV transport update in a $(D\!-\!1)$-dimensional fracture grid. 
In this work, we combine the implicit MHFE transport and pressure update and the CFE discrete fracture model with an \textit{implicit} higher-order DG transport update, both in the fractures and the matrix. We demonstrate that with this approach the CFE method has low numerical dispersion and even higher computational efficiency than in \citet{zidane2014efficient}. A wide range of test cases demonstrate that this all-implicit method can achieve the same numerical accuracy as the IMPEC scheme, but with up to 5000 times larger time-steps. This method is also for the first time applied to 2D and 3D unstructured grids (using triangular, quadrilateral, and hexahedral elements). Because we sequentially use an implicit update for the pressures and fluxes (MHFE) and an implicit transport update (DG), we refer to this scheme as implicit-pressure-implicit-compositions, or IMPIC, and the discrete fracture model as IMPIC-CFE (to distinguish it from a fully coupled implicit scheme).

Our proposed method only requires the inversion of one global matrix for the higher-order DG transport update, regardless of the number of components. Furthermore, by using efficient sparse matrix solvers, the CPU time for each implicit update is only 2--3 times more than one explicit update (while requiring orders of magnitude fewer time-steps). Together, this results in a fast higher-order method for discretely fractured reservoirs. 
{This sets the stage to accurately and efficiently study a wide range of large-scale problems in hydrogeology and hydrocarbon reservoirs for which high numerical dispersion is undesirable (e.g., viscous and gravitational flow instabilities), or for which homogenized fracture representations (e.g., dual-porosity/permeability models) or limiting assumptions on the physics (e.g., incompressibility, number of species, or correlations for fluid properties instead of an equation-of-state) are overly restrictive. One hydrogeological application is carbon sequestration in fractured saline aquifers. To model the effectiveness and measurements of CO$_{2}$ solubility trapping, we may also need to consider other dissolved species, such as (co-injected or natural) methane, different isotopes (which can be modeled as separate species), and noble gasses (when used as tracers of CO$_{2}$ migration). Some of these species (e.g., methane and salts) may affect the CO$_{2}$ solubility and phase behavior of the brine. The compressibility of brine (and even the formation) will determine the pressure response to CO$_{2}$ injection. A density increase of the aqueous phase upon CO$_{2}$ dissolution, as predicted by an EOS, could trigger (the small-scale onset of) density driven flow, which requires exceedingly fine grids (when using lowest-order methods) or higher-order methods to resolve. And finally, the location and orientation of a few large fractures could determine whether injected CO$_{2}$ will remain trapped within the intended aquifer extent.}

In the following, we summarize the mathematical description of compressible single-phase multicomponent flow and present our numerical models for IMPEC and IMPIC discrete fracture simulations using the combination of MHFE with higher-order DG. A large number of numerical experiments investigate the convergence rates of different combinations of temporal and spatial discretizations. We consider discrete fracture networks in 2D and 3D {irregular} grids, and, finally, we test the robustness of the approach for a highly non-linear gravitational fingering problem on {irregular}, and one fractured, grids. Comparisons with a commercial reservoir simulator are made for both fractured and unfractured examples.

\section{Mathematical Model}\label{sec::mathmodel}
Compressible single-phase flow of a $n_{c}$ component fluid is governed by transport (mass conservation) equations for each species $i$, Darcy's law relating the advective flux $\vec{v}$ to gradients in pressure ($p$) and gravitational acceleration ($\vec{g} = g \vec{z}$, with $\vec{z} = (0, 0, z)$ pointing up), and a pressure equation from volume balance \citep{acs, watts}:
\begin{eqnarray}\label{eq::eq1}
&&\phi \frac{\partial c_{i}}{\partial t} + \nabla\cdot (c_{i} \vec{v})  =  F_{i} \quad \quad \forall\ i=1, \ldots, n_{c},\\\label{eq::eq2}
&&\vec{v}  =  - \frac{\mathrm{K}}{\mu} \nabla (p - \rho  g z), \\\label{eq::eq3}
&&\phi C_{f} \frac{\partial p}{\partial t} +  \sum_{i=1}^{n_{c}} \bar{\nu}_{i} \left[\nabla\cdot (c_{i} \vec{v}) -F_{i}\right]  =  0,
\end{eqnarray}
in terms of rock properties: porosity $\phi$ and absolute permeability tensor $\mathrm{K}$. The fluid is described by the molar density $c_{i}$ of species $i$, composition dependent viscosity $\mu$ and mass density $\rho$, partial molar volumes $\bar{\nu}_{i}$, and fluid compressibility $C_{f}$. $F_{i}$ represents sink and source terms, such as contaminant spill sites in hydrogeological applications, or injection and production wells in hydrocarbon reservoir modeling. 

Eqs.~(\ref{eq::eq1})--(\ref{eq::eq3}) are $n_{c} +4$ equations for the $n_{c}+4$ unknowns $p$, $v_{x}$, $v_{y}$, $v_{z}$, and $c_{i}$. The molar density of the mixture is $c=\sum_{i} c_{i}$. Molar fractions of each component can be defined as $z_{i} = c_{i}/c$. The mass density is related to the molar density through the molecular weights of each species $\mathrm{MW}_{i}$ as $\rho = \sum_{i} (c_{i} \mathrm{MW}_{i}$). The partial molar volumes and fluid compressibility are temperature, pressure, and composition dependent, and are derived rigorously from an equation-of-state (EOS) \citep{moortgatIII}. Specifically, we use the Cubic-Plus-Association (CPA; \citet{liCPA}) EOS when considering an aqueous phase. The CPA EOS reduces to the Peng-Robinson (PR; \citet{preos}) EOS for hydrocarbon phases. Fluid viscosities are derived from the LBC method \citep{viscosity} or corresponding states model \citep{viscosity2}.

Fluid flow and transport in fractures, while discretized differently, is described by the same Eqs.~(\ref{eq::eq1})--(\ref{eq::eq3}), with the porosity taken as one, and the intrinsic fracture permeability $\mathrm{K}$ often derived from the fracture aperture.

\section{Numerical Methods}\label{sec::numerical}
Eqs.~(\ref{eq::eq1})--(\ref{eq::eq3}) are non-linearly coupled. The most robust solution method would solve all $n_{c}+4$ equations implicitly (i.e., simultaneously) over the entire 3D grid, using, for instance, the Newton-Raphson (NR) method. Even in that approach, additional non-linearities, such as the composition dependence of fluid compressibility, partial molar volumes, and viscosities, are generally updated explicitly within each NR iteration. While robust, the fully implicit method requires the inversion of very large matrices with non-trivial sparsity patterns, and as such is also the most computationally expensive.  

To achieve a higher efficiency, at the expense of some accuracy, one can decouple
Eq.~(\ref{eq::eq1}) from Eqs.~(\ref{eq::eq2})--(\ref{eq::eq3}). First, pressures and fluxes are updated using fluid properties ($c_{i}$ in this case) from the previous time-step, and then the transport equation is updated in terms of a given velocity field. This leads to the common implicit-pressure-explicit-composition (IMPEC) scheme, and the implicit-pressure-implicit-composition (IMPIC) approach that we adopt in this work. 
Additional iterations between the two updates can reduce the decoupling error, but increases the computational cost.

Apart from resulting in two decoupled problems that are more numerically tractable, a benefit of the IMPEC/IMPIC approach is that each problem can be solved with a different appropriate numerical method. In this work, we adopt the MHFE method to, simultaneously and to the same order, solve for the pressure and velocity fields. MHFE allows for any permeability tensor, unstructured grids, and is suitable for discrete fracture implementations (e.g, \citet{hoteit2005, hoteit2008}). The transport update is discretized by the DG method, which allows for sharp discontinuities in fluid properties at the fracture-matrix interface.
At lowest order and using upwind numerical fluxes the DG method reduces to the {single-point upstream weighted lowest-order} finite volume (FV) method. By using a multilinear DG approximation, numerical dispersion can be reduced significantly. We implement both FV and DG and compare the performance in the Numerical Experiments. For the IMPEC and IMPIC schemes we use the forward and backward Euler discretizations. A second-order Crank-Nicolson scheme is discussed as well.

Fractures are modeled with the cross-flow-equilibrium model that was first presented in 2D by \citet{hoteit2005} and generalized to 3D by \citet{moortgatVI}. In the CFE model, fractures are combined with a small slice of matrix on either side into larger $D$-dimensional grid cells in the same mesh as the matrix. The flux contributions from both the fracture and the matrix portions of the CFE elements are integrated on any unstructured grid type by the MHFE method. Once the fluxes are obtained, the transport update is identical to that on an unstructured grid. Because we only consider new numerical methods for the transport update, the fracture model will not be discussed further and we refer to \citet{hoteit2005, moortgatV, moortgatVI} for more details.

In the following, we first consider the lowest-order FV discretization and provide a unified framework for all the considered time-discretizations. Next, we present explicit and implicit second-order DG transport updates.

\subsection{Explicit and Implicit Finite Volume Methods for Multicomponent Flow}
The FV discretization of the mass balance equation is well-known. Assuming element-wise-constant values of the state variables, molar densities $c_{i,K}$, and edge-wise-constant normal components of fluxes $q_{K, E \in \partial K}$, for each grid cell $K$ and edge $E$, the weak form of the transport equations is:
\begin{equation}\label{eq::tranport}
c_{i, K}^{n+1} = c_{i, K}^{n} - \frac{\Delta t}{\phi_{K} V_{K}}\left( \sum_{E\in\partial K}  q_{K,E} \widetilde{c_{i, K}}^{m} - V_{K} F_{i,K} \right)
\end{equation}
with $\phi_{K} $ and $V_{K}$ the porosity and volume of grid cell $K$, and $F_{i,K}$ a source/sink in $K$. Communication between grid cells is facilitated through the upwinding of $c_{i}$ across edges: $\widetilde{c_{i, K}} = c_{i,K}$ if $q_{K,E}\ge 0$ and $\widetilde{c_{i, K}} = c_{i, K^{\prime}}$ if $q_{K,E}< 0$, with $K^{\prime}$ the element neighboring edge $E$.
The superscript $n$ denotes the time-step, and Eq.~(\ref{eq::tranport}) is written such that for $m=n$ the temporal discretization is the explicit forward Euler method, while for $m=n+1$ we obtain the implicit backward Euler approximation. 

For the explicit forward Euler method, Eq.~(\ref{eq::tranport}) can be easily implemented in a loop over all grid cells $K$. Alternatively, Eq.~(\ref{eq::tranport}) can be solved using matrix algebra, which is useful, for instance, in the context of interpreter languages (e.g., the open source MRST, Matlab Reservoir Simulation Toolbox \citep{mrst}). We define the identity matrix $\mathbf{I}$, and the diagonal matrix $\mathbf{dt}$ with elements $[\mathrm{dt}]_{K,K^{\prime}} = [\mathrm{I}]_{K,K^{\prime}} \Delta t/(\phi_{K} V_{K})$, and $\mathbf{f_{i}}$ with $[\mathrm{f}_{i}]_{K,K^{\prime}} = [\mathrm{I}]_{K,K^{\prime}} \Delta t F_{i,K}/\phi_{K}$. Eq.~(\ref{eq::tranport}), with $m=n$, then becomes
\begin{equation}\label{eq::transportMatrixform}
\mathbf{c}_{i}^{n+1} = \mathbf{A} \mathbf{c}_{i}^{n} + \mathbf{f_{i}} , \quad\mbox{with}\quad \mathbf{A} =  \left( \mathbf{I} - \alpha \mathbf{dt} \mathbf{Q}\right),
\end{equation}
with $\alpha = 1$. The matrix $\mathbf{Q}$, when multiplied with a vector $\mathbf{c}_{i}$ of molar densities $c_{i,K}$ in each grid-cell $K$, provides the sum of upwind molar fluxes entering and leaving each grid-cell. The elements of $\mathbf{Q}$ are given in \citet{mrst},

A benefit of re-writing Eq.~(\ref{eq::tranport}) as Eq.~(\ref{eq::transportMatrixform}) is that we also obtain an expression for the \textit{implicit} backward Euler update by setting $m=n+1$, $\alpha = -1$:
\begin{equation}\label{eq::transportMatrixformImpl}
\mathbf{c}_{i}^{n+1} = \mathbf{A}^{-1} (\mathbf{c}_{i}^{n} + \mathbf{f_{i}}).
\end{equation}

Finally, we can write the second-order Crank-Nicolson scheme as
\begin{equation}\label{eq::transportMatrixformCN}
\mathbf{c}_{i}^{n+1} = \mathbf{A}_{-}^{-1}  (\mathbf{A}_{+}\mathbf{c}_{i}^{n} - \alpha \mathbf{f_{i}}) - \alpha \mathbf{f_{i}},
\end{equation}
with $\mathbf{A}_{\pm}$ denoting $\alpha = \pm 1/2$.

It is important to note that due to the decoupling of the flow and transport equations, Eqs.~(\ref{eq::transportMatrixform})--(\ref{eq::transportMatrixformCN}) are \textit{linear} in $\mathbf{c}_{i}$ (for a given flux field), and can thus be solved by direct solvers, i.e.~without the need for iterative non-linear methods, such as Newton-Raphson. Moreover, the matrix $\mathbf{Q}$ for fluxes, and therefore also $\mathbf{A}$, does not depend on compositions! The consequence is that, regardless of the number of components, only one global matrix $\mathbf{A}$ needs to be inverted for the implicit update. Eqs.~(\ref{eq::transportMatrixform})--(\ref{eq::transportMatrixformCN}) provide highly efficient (though possibly dispersive) schemes for all three time-discretizations. 

\subsection{Explicit and Implicit Second-Order Discontinuous Galerkin Methods}
For the DG transport update, we can write, similar to Eq.~(\ref{eq::tranport})
\begin{eqnarray}\label{eq::dgupdate} 
&& c^{n+1}_{i,K,N} = c^{n}_{i,K,N} - \frac{\Delta t}{\phi V_{K}}  [F(c^m_{i,K,N},q_{E}) - G(\widetilde{c_{i,K,N}^m},q_{E}) - V_{K} F_{i,K}].
\end{eqnarray}
Three important differences are:
\begin{enumerate}
\item each grid-cell $K$ now has $N_{N}$ degrees of freedom at the $N_{N}$ vertices, e.g.~3 for triangles, 4 for quadrilaterals, and 8 for hexahedra,
\item the weak form of the transport equation in the DG discretization uses a partial integration that results in one volume integral that involves only the variables inside the grid-cell, $F(c_i^m,q_{E})$, and one surface integral in which edge-fluxes are multiplied with values from the upwind direction, $G(\tilde{c}_i^m,q_{E})$, which may be in neighboring grid-cells,
\item \textit{multiple} nodal values are upwinded for each face (2 in 2D; 4 in 3D).
\end{enumerate}

The implications are that the DG update of \textit{each} individual node $N_{N}$ involves up to 4 nodes (for hexahedra) inside element $K$, and, depending on the flux directions, up to 3 (upwind) nodes in 3 neighboring grid cells that share the same node. Combined, up to 32 local node values are involved in the update of each hexahedral grid cell $K$ (8 nodes in $K$, plus 4 upwind nodes for each of the 6 neighbors), 12 for quadrilaterals (4 nodes plus $2$ per 4 edges), and 9 for triangles. Detailed expressions for the element-by-element updates on triangular, quadrilateral, prismatic, tetrahedral, and hexahedral grids are provided in \citet{moortgatIX}.

The resulting global matrix system for the explicit DG update is again of the form (see Appendix):
\begin{equation}\label{eq::transportMatrixformDG}
\mathbf{\bar{c}}_{i}^{n+1} = \mathbf{\bar{A}} \mathbf{\bar{c}}_{i}^{n} + \mathbf{f_{i}} , \quad\mbox{with}\quad \mathbf{\bar{A}} =  \left( \mathbf{I} - \mathbf{dt} \mathbf{\bar{Q}}\right), 
\end{equation}
but the vector $\mathbf{\bar{c}}_{i}$ is now $N_{N}\times N_{K}$ long, and the matrices $\mathbf{\bar{A}}$, $\mathbf{\bar{Q}}$ (and $\mathbf{I}$) all $N_{N}^{2}\times N_{K}^{2}$, with $N_{K}$ the total number of grid-cells in the domain. Bars (e.g.~$\mathbf{\bar{c}}_{i}$) indicate that values are different at each node, while $\mathbf{I}$, $\mathbf{dt}$, and $\mathbf{f_{i}}$ have the same value at each node for a given element $K$. 

Once the matrix $\mathbf{\bar{A}} $ is constructed, the implicit backward Euler and Crank-Nicolson methods can again be elegantly written as in Eqs.~(\ref{eq::transportMatrixformImpl})-(\ref{eq::transportMatrixformCN}).

It is clear that the DG update results in a larger matrix system than required for the FV method. In the Numerical Experiments section we demonstrate that this is compensated for by the higher-order convergence rate achieved by the DG method. For any specific desired accuracy in the results, the DG method can be considerably faster than the FV method, because much coarser grids can be used.

{Sparsity patterns for $\mathbf{A}$ and $\mathbf{\bar{A}}$ are shown in Figure~\ref{fig::dofquad} for different flux fields, determined by well locations and gravity. Interesting recent work (e.g., \citet{kwok2007potential, natvig2008fast}) demonstrates that further efficiency gains can be obtained by re-ordering the grids cells in relation to the global flux (or alternatively, pressure) fields. Such optimizations of our IMPIC schemes could be the subject of future work.}

\section{Numerical Experiments}
In this section, we demonstrate the performance of the new implicit numerical methods developed in this work. The first example considers first-contact-miscible (FCM) \co injection into a multicomponent compressible oil, simulated on {irregular} 2D and 3D (unfractured) grids. The purpose of this example is to investigate and quantify the computation efficiency versus the numerical accuracy (convergence rate) for the different numerical methods considered in this paper: 
\begin{enumerate}
\item spatial: lowest-order finite volume (FV) {with single-point upstream weighting} and discontinuous Galerkin (DG), both in combination with MHFE for pressures and fluxes,
\item temporal: lowest-order forward (explicit) and backward (implicit) Euler, as well as second-order implicit Crank-Nicolson transport updates, each combined with implicit MHFE (e.g.~IMPEC, and IMPIC)
\end{enumerate}

In the second example, we discuss the most powerful application of the higher-order implicit transport models: discretely fractured reservoirs. We also compare our CFE model to a single-continuum discretization with a FV commercial reservoir simulator. The last example considers the detrimental impact of numerical dispersion when simulating complex flow problems, in this case gravitational instabilities, that require high resolution to resolve. This example is compared to a commercial reservoir simulator as well.

{To clarify, all grid types are implemented as unstructured finite elements that allow for non-Cartesian ordering. However, to facilitate comparisons to structured grids, the quadrilateral and hexahedral examples consider logically-Cartesian grids. We adopt the term `{irregular}' grids to refer to elements with non-orthogonal faces.} 

\subsection{Example 1: Numerical Efficiency, Accuracy, and Convergence of FV and DG Implicit Transport Updates on 2D and 3D {Irregular} Grids}
To study the convergence properties of the numerical methods developed in this work, we model FCM \co injection with gravity into $100\ \times 100\ \mathrm{m}^{2}$ and $100\ \times 100\ \times 100\ \mathrm{m}^{3}$ two- and three-dimensional domains, respectively. The domain is initially saturated with a 5 (pseudo)-component oil. The compositions and critical properties of the oil are given in Table~\ref{table::FCMcomp}. Table~\ref{table::conditions} provides the density and viscosity of both oil and injected \coo, as well as a 50 mol\% to 50 mol\% mixture. At the reservoir temperature and pressure (Table~\ref{table::conditions}), \co is considerably denser than the oil, so injection is from the bottom-left corner at a constant rate of 5\% pore volume (PV) per year. Production is at constant pressure from the opposite corner. The porosity is 20\% and the homogeneous and isotropic permeability is 10 md.

Convergence is analyzed on three grid types: triangular, quadrilateral (2D), and hexahedral (3D). For this analysis, the latter two are logically Cartesian, but grid nodes are randomly perturbed by up to 30\% of the grid size, resulting in the poor quality elements that are generally required to conform to realistic geological formations. {Grid sizes (and porosities) are still relatively uniform in this example to allow reasonable time-step sizes for the IMPEC method. Domains with a wide range in grid sizes (due to fractures) become prohibitively CPU expensive in IMPEC and require implicit methods, as will be demonstrated in the next examples.}

Coarse grid simulations are compared to a reference solution on a finest grid. Specifically, in 2D, the reference solution is an IMPEC-DG simulation on a $320\ \times \ 320$ quadrilateral grid. In 3D, the reference is an IMPEC-DG simulation on a $50\ \times\ 50\ \times 50$ element hexahedral grid. To obtain convergence rates, we compute the \error error over the entire domain.

\begin{figure}[h!]
 \centering
\includegraphics[width=\textwidth]{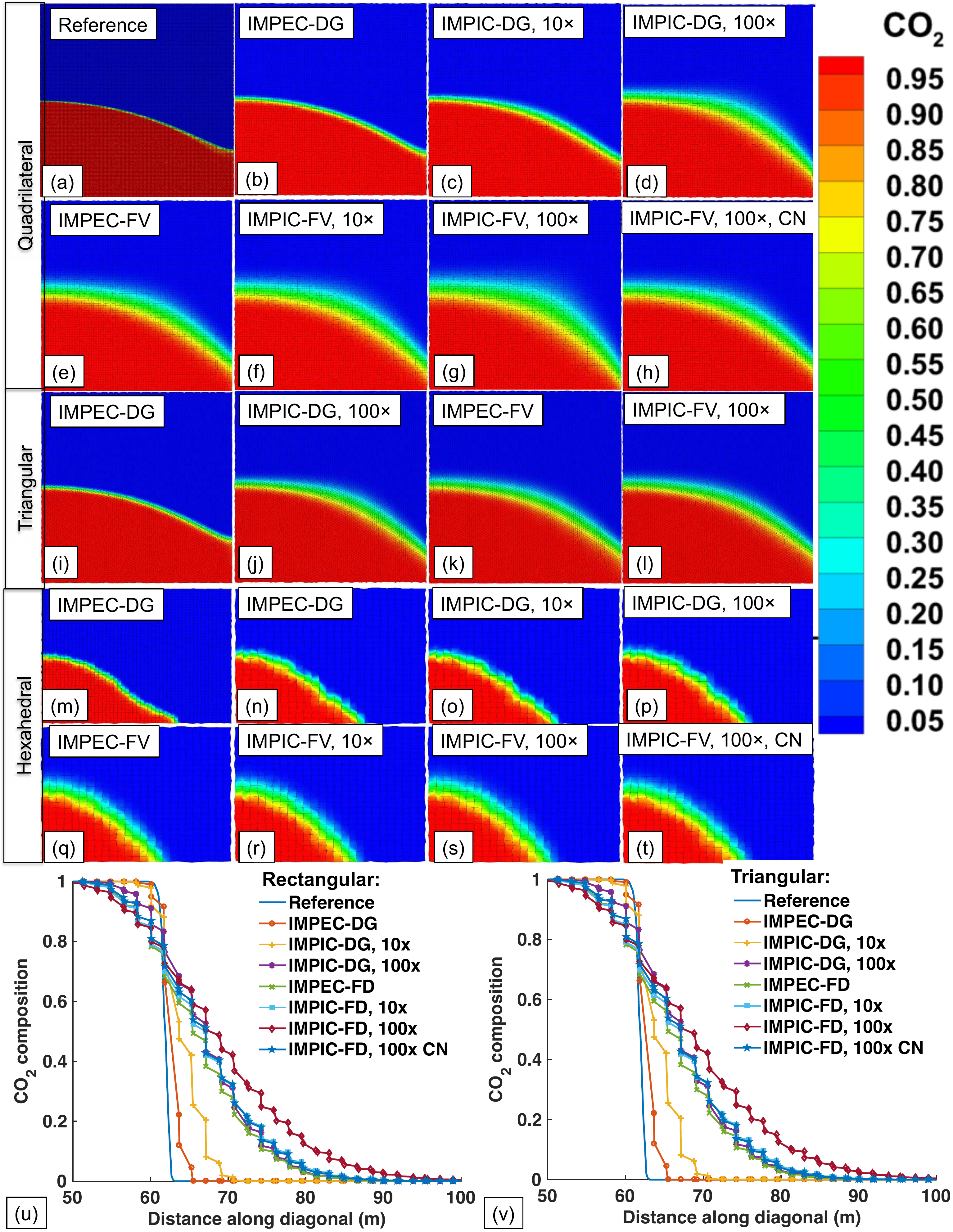} 
\caption{\co molar fraction 40\% PVI on quadrilateral \textbf{(a)}--\textbf{(h)} and triangular \textbf{(i)}--\textbf{(l)} grids and at 15\% for a diagonal vertical cross-section injector to producer for hexahedral grids \textbf{(m)}--\textbf{(t)}. IMPEC-DG on $320\times320$ grid \textbf{(a)} and $50\times50\times50$ \textbf{(m)} grids are the reference; other results are shown for $80\times80$ quadrilaterals \textbf{(b)}--\textbf{(h)}, triangles with $h=1.25\ \mathrm{m}$ \textbf{(i)}--\textbf{(l)}, and a $25\times25\times25$ grid \textbf{(n)}--\textbf{(t)}. \co molar fraction on a diagonal line from injector to producer are shown for all quadrilateral \textbf{(u)} and triangular \textbf{(v)} grids.}
 \label{fig::ex1a}
 \end{figure}

For each grid type, simulations are performed on different levels of mesh refinement with the characteristic element size $h$ computed as $\sqrt{2V}$ for triangles, $\sqrt{V}$ for quadrilaterals, and $\sqrt[3]{V}$ for hexahedra, with $V$ the average element volume. All simulations use the implicit MHFE method to compute pressures and fluxes, and the focus is on different explicit and implicit transport updates.
Convergence properties and computational times are studied for both DG and FV in combination with: 1) IMPEC, 2) the implicit-pressure-implicit-composition (IMPIC) scheme, but with time-steps equal to IMPEC, i.e., constrained by the CFL condition, 3) IMPIC with 10 times larger time-steps ($10\times$CFL), 4) IMPIC with $100\times$CFL, 5) IMPIC with $1000\times$CFL, and 6) IMPIC with $100\times$CFL but using a Crank-Nicolson second-order time discretization. Together, around 200 simulations were carried out for this example, run in serial on a 2.8GHz Intel i7 processor. The results are summarized in Figures~\ref{fig::ex1a} and \ref{fig::ex1b}.

Figure~\ref{fig::ex1a} provides snap-shots of the \co molar fraction throughout the domain, {as well as 1D cross-sections}, to visually illustrate the (lack of) numerical dispersion on the reference grid and on an intermediately coarser grid for different numerical methods and different grid types. Figures~\ref{fig::ex1a}b--c show that on a $80\times 80$ quadrilateral grid IMPEC-DG and IMPIC-DG with time-steps up to $10\times$CFL have low numerical dispersion as compared to the reference simulation on a $320\times320$ grid, while all FV simulations show lower accuracy. On triangular grids with $h=1.25$, IMPEC-DG results are close to the reference, while IMPEC-DG with $100\times$CFL shows comparable dispersion to an IMPEC-FV simulation.  3D results are shown for $50\times 50\times 50$ and $25\times 25\times 25$ grids for a vertical cross-section that runs diagonally from the injector at (0, 0, 0) and the producer at (100 m, 100 m, 100 m). Compared to the 3D reference solution, the subsequent panels (Figures~\ref{fig::ex1a}n--t) show increasing dispersion when moving from IMPEC-DG to IMPIC-DG with larger time-steps, then IMPEC-FV and IMPIC-FV with larger time-steps. A Crank-Nicolson scheme does not significantly reduce the numerical dispersion in the FV results. 

To provide a more quantitative analysis, the \error errors are plotted logarithmically versus the grid sizes $h$ in Figure~\ref{fig::ex1b}, such that the slopes provide the convergence rates. The errors are not normalized, and show the average deviation in \co composition with respect to the reference solution. Auxiliary curves illustrate linear and quadratic convergence rates. 

As expected, IMPEC-DG has the lowest numerical dispersion and highest convergence rate. On relatively fine grids, the convergence rate is nearly quadratic, but on the coarsest grids the slope decreases. 
IMPIC-DG simulations with the same time-step size as IMPEC show only slightly more dispersion. However, when the time-step sizes are increased 10, 100, and 1000-fold, the \error errors increase and convergence rates decrease, because of the linear convergence of the Euler method. Still, IMPIC-DG simulations with $100\times$CFL time-steps have the same accuracy as FV simulations with IMPEC or IMPIC with $1$--$10\times$CFL. When the Crank-Nicolson scheme is used, which is of the same order as the DG spatial discretization, IMPIC-DG simulations with $100\times$CFL have considerably less dispersion, close to that of IMPIC-DG Euler results with 10 times smaller time-steps. 

For the FV simulations, the formal convergence rate for both the spatial and temporal discretizations are linear and the impact of using IMPIC with increasingly large time-steps (with $\Delta t\propto h$) is less pronounced as compared to IMPEC. All FV simulations have considerably more numerical dispersion than all higher-order DG simulations. While both FV and DG convergence rates are below the formal linear and quadratic orders, respectively, the DG convergence rate is still consistently about twice that of FV. The FV \error errors also have a higher overall multiplicative factor with respect to $h$, which accounts for the upward shift of the FV error plots in Figure~\ref{fig::ex1b}, in addition to the lower convergence rate (slope). The comparison between IMPEC-DG and IMPIC-FV is in line with earlier convergence analyses on structured grids in \citet{moortgatIII, moortgatIX}.

The most powerful way to visualize the impact of convergence rates (or, equivalently, numerical dispersion) is to plot the computational (CPU) times for each simulation versus the \error errors (Figure~\ref{fig::ex1b}). Comparing the computational cost of different numerical methods on the same grid is meaningless when they produce different results. A more appropriate way to evaluate the performance of different methods is to compare their computational costs for a given numerical accuracy. Without doing this, it is not obvious whether, say, an IMPIC-DG simulation is more efficient than an IMPEC-DG simulation, because 1) IMPIC allows for larger time-steps, but 2) each time-step requires more computational cost, and 3) a finer mesh may be required to reduce the numerical dispersion from IMPIC with large time-steps. 

Figure~\ref{fig::ex1b} demonstrates that for relatively uniform grids and using explicit and implicit Euler updates, IMPEC-DG is in fact more computationally efficient than all other methods, due to its low numerical dispersion and small computational cost per (albeit small) time-step. IMPIC-DG with the same time-step size has only slightly worse accuracy, but is more computationally expensive per time-step, and as such requires about twice the CPU time to achieve the same accuracy as IMPEC-DG. IMPIC-DG with $10\times$CFL has more numerical dispersion, but requires roughly 10 times less computational time, allowing for finer grid simulations at the same cost as a coarser grid IMPIC-DG with $1\times$CFL, which is why the curves are close to each other. When the (backward Euler) time-step size is increased further, the computational efficiency decreases due to the need for mesh refinement to reduce numerical dispersion. IMPIC-DG with $10\times$CFL is still more efficient at each level of accuracy than any IMPEC or IMPIC-FV method. When the second-order Crank-Nicolson method is combined with the second-order IMPIC-DG, we see the best performance, because we can use large time-steps without the first-order increase in numerical dispersion from the Euler method. 

All lowest-order methods require considerably more CPU time than the higher-order methods to achieve a given accuracy, because much finer grids have to be used to achieve the same results. For FV simulations, though, the computational efficiency \textit{does} increase (relative to IMPEC) by using IMPIC with larger time-steps. This is because the IMPEC-FV spatial discretization error is already lowest-order (linear) in $h$, so the rate of convergence is not reduced as much by the Euler discretization error which is linear in $\Delta t \propto h$. Figure~\ref{fig::ex1b} shows that IMPEC-FV simulations already have such high numerical dispersion that the IMPIC-FV with $100\times$CFL time-steps is only slightly worse. However, the latter is about a hundred times faster, and as such IMPIC-FV can be more efficient than IMPEC-FV. 

The above discussion mostly applies to each grid type. However, on fine 3D grids the matrices that need to be inverted in the IMPIC-DG (and to a lesser extent IMPIC-FV) update become extremely large and the direct solver used for this analysis (Pardiso, included in the Intel Math Kernel Library) becomes inefficient. Parallel iterative solvers should be used to further improve the computational efficiency of the implicit methods. 

{We also note that the grids in this example have close to unit aspect ratios. We performed additional IMPEC and IMPIC simulations for $25\times 256$ quadrilateral grids with high aspect ratios, and find that for such grids we observe no grid sensitivity in terms of preferential flow directions, but that numerical dispersion is of course higher in the coarse $x$-direction than in the 10 times finer $y$-direction (not shown). A more detailed study of grid sensitivity is performed in \citet{moortgatIX}.}

The conclusions from this example are that 1) the higher-order DG methods are two to three orders of magnitude faster than lower-order FV ones, and 2) on grids with relatively uniform grid sizes, the IMPEC-DG scheme is actually more numerically efficient than IMPIC-DG, unless a Crank-Nicolson scheme is used with very large time-steps. 

For discretely fractured reservoirs, though, or any other grid with a low number of very small elements (or low porosities), the CFL time-step constraint for the entire domain may be determined by those few small grid cells (unless an adaptive time-stepping method is used), resulting in poor performance of the IMPEC scheme. In such applications, the unconditional stability of the IMPIC scheme can significantly outperform IMPEC, as we discuss in the next examples for fractured reservoirs.

\subsection{Example 2: C$_{1}$ into C$_{3}$ Injection on Structured 2D Grid with 12 Discrete Fractures}\label{sec::ex2}
To test our higher-order implicit FE methods on discretely fractured domains, we first compare to an example from the literature \citep{zidane2014efficient} { and also model the same problem with a commercial reservoir simulator.} We consider a $500\times 200\ \mathrm{m}^{2}$ structured 2D grid with 6 horizontal and 6 vertical discrete fractures, all with a permeability of $10^{3}$ d and an aperture of $0.1$ mm. CFE fracture elements have a width of 30 cm. The matrix porosity and permeability are 20\% and 1 md, respectively.
Methane (C$_{1}$) is injected from the bottom-left corner at a high rate of 50\% PV/yr, constant pressure production is from the top-right corner, and the domain is initially saturated with propane (C$_{3}$). Fluid densities and viscosities and the reservoir temperature and initial pressure are given in Table~\ref{table::conditions}.

\begin{figure}[h!]
 \centering
\includegraphics[width=\textwidth]{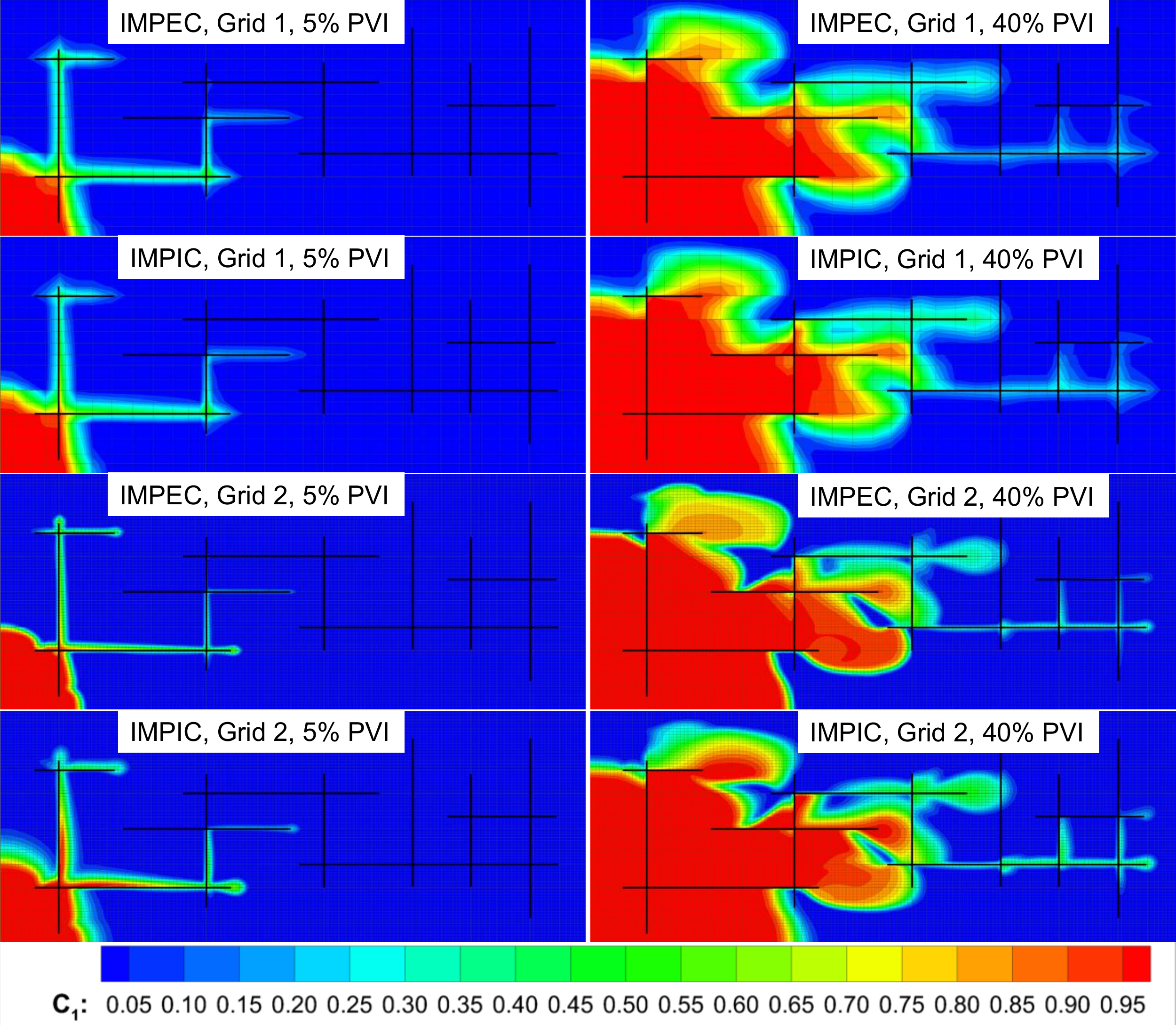} 
\caption{Example 2: Methane (C$_{1}$) molar fraction at 5\% (left column) and 40\% (right column) PVI. C$_{1}$ is injected at 50\% PV/yr from the bottom-left corner into propane, with constant pressure production from the top-right corner.
Grid 1 has $46\times 26$ elements, Grid 2 has $166\times 86$. Results are shown for Implicit-Pressure-\textbf{Explicit}-Concentration (IMPEC) and Implicit-Pressure-\textbf{Implicit}-Concentration (IMPIC) MHFE-DG simulations.
Discrete fracture locations (12) are illustrated with an exaggerated thicknesses.}
 \label{fig::fracsC1C3}
 \end{figure}

Figure~\ref{fig::fracsC1C3} shows the domain, the fracture locations, and the overall molar fraction of methane throughout the domain at 5\% and 40\% PVI on both a coarse Grid 1 with $86\times 46$ elements and a finer $166\times 86$ element Grid 2. We compare the accuracy and computational cost of our IMPEC and all-implicit IMPIC schemes. For the latter we set the time-step size as $1000$ and $5000$ times the (adaptive) CFL condition for Grids 1 and 2, respectively. The total CPU time, total number of time-steps, and CPU time per time-step for all the numerical examples in this section are provided in Table~\ref{table::ex4-performance}.

Figure~\ref{fig::fracsC1C3} shows that both implicit and explicit MHFE-DG with the CFE fracture model have the same accuracy, or level of numerical dispersion, on the coarsest Grid 1 as the results presented in \citet{zidane2014efficient}. The results also agree reasonably well with those on the finer Grid 2. In terms of CPU efficiency, of course we find that the IMPEC simulations with narrow CFE elements require many small time-steps. However, for our higher-order IMPIC method the CPU time actually appears to be less than for the method with a FV implicit fracture update in \citet{zidane2014efficient}. 

Promisingly, while our IMPIC method requires three orders of magnitude fewer time-steps than IMPEC, each implicit DG transport update for the entire grid requires only $2$--$3$ times the CPU time of a single explicit update. 

\begin{figure}[h!]
 \centering
\includegraphics[width=\textwidth]{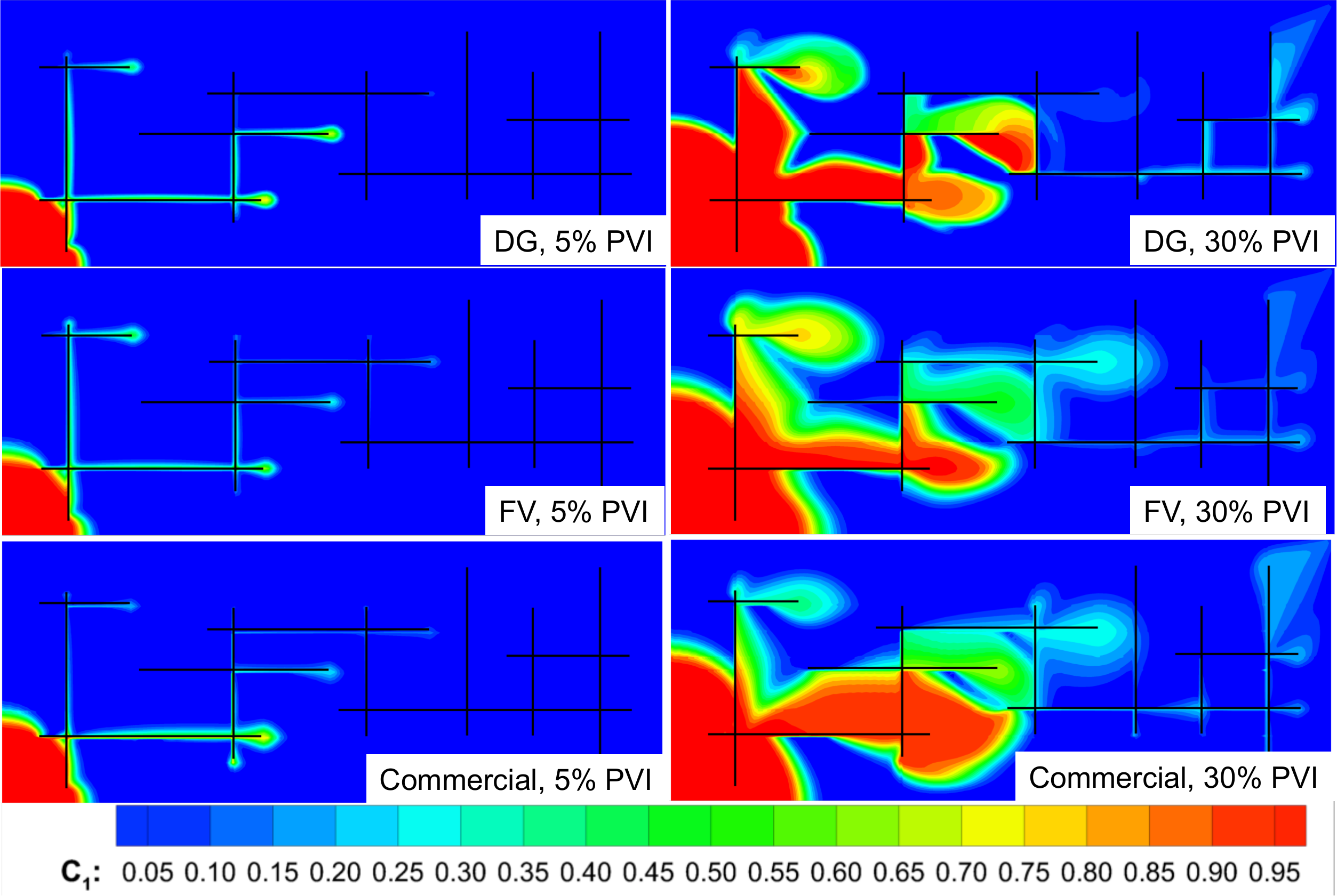} 
\caption{Example 2: Methane (C$_{1}$) molar fraction at 5\% (left column) and 30\% (right column) PVI (at 5\% PV/yr). Results are shown for IMPIC-DG (top row) and IMPIC-FV (middle row) simulations with 1 mm wide CFE elements and for a fully implicit FV simulation with a commercial reservoir simulator (bottom row) with 0.1 mm wide explicitly discretized fracture grids cells.}
 \label{fig::fracsC1C3b}
 \end{figure}

{We also carry out a single-continuum simulation with the fully implicit option of a FV commercial reservoir simulator. In this simulation, fracture grid cells are discretized explicitly by 0.1 mm wide grid cells, which are given unit porosity and the fracture permeability of 1 d. For this comparison, we inject at 5\% PV/yr and compare to IMPIC-DG and IMPIC-FV simulations with CFE elements of 1 mm width. Note that to compare to IMPEC simulations, the CFE elements have to be given relatively large sizes to allow reasonable CFL time-step sizes, but IMPIC has no such restrictions. 

Figure~\ref{fig::fracsC1C3b} shows the results for the methane composition after one and six years. As discussed in the Introduction, implicit updates of single-continuum fracture discretizations can result in ill-conditioned matrix inversions. This was indeed observed in the commercial reservoir simulator, which failed to fully converge in 70\% of the time-steps, despite (internally) cutting the average time-step size to 5.5 hrs and manually tuning the maximum number of iterations. Our IMPIC CFE simulations allowed for time-step sizes of 2 weeks and only required 45 and 89 secs for IMPIC-FV and IMPIC-DG, respectively, as compared to almost two hrs for the commercial reservoir simulator. Despite the numerical issues in the single-continuum implicit FV simulation, the results are quite similar to our IMPIC-FV method, but the latter shows two orders of magnitude higher efficiency. Further improvements are provided by IMPIC-DG which has sharper fronts (less numerical dispersion) at comparable computational cost. }

\subsection{Example 3: C$_{1}$ into C$_{3}$ Injection on {Irregular} 2D and 3D Fractured Grids}
An attractive feature of our CFE model is that the transport update in the fracture (CFE) elements is identical to those in the matrix. As such, the implicit FV and DG transport updates presented in this work automatically carry over to the unstructured grids for which the explicit methods were initially developed.  
We generalize the results from the previous test case to non-orthogonal 2D and 3D grids. Apart from the grids, all simulation parameters are the same as before. For the 2D grid (referred to as Grid 3, and shown in Figure~\ref{fig::fracsC1C3FD}a), we modify the nodal coordinates of the fine Grid 2 in the $y$-direction sinusoidally to create a dome-like structure and add a 10\% incline. The $x$-coordinates are also `stretched' wider in the upward direction to create an {irregular} quadrilateral grid. The fractures are kept in the same logically-Cartesian locations. For the 3D grid (Grid 4), we extend the aforementioned  2D grid by 50 m (5 elements) in the vertical direction and add another sinusoidal variation in $z$ with respect to $x$. The 12 previous fractures extend vertically from the bottom to the top of the domain, and an additional horizontal fracture is placed in the middle, as illustrated in Figure~\ref{fig::fracsC1C3FD}b.
MHFE-DG simulations for methane injection are performed on both grids with our implicit transport update (IMPIC). The time-step sizes are $5000$ times the CFL constraint.

Figures~\ref{fig::fracsC1C3FD}c-d show that the methane concentration profiles at 5\% and 40\% PVI, apart from the different geometry, are quite similar to those on the structured 2D Grid 2 (Figure~\ref{fig::fracsC1C3}), with no signs of mesh sensitivity or numerical artifacts due to the {irregular}ity of the grids. 
The 3D results in Figures~\ref{fig::fracsC1C3FD}e-f are also initially similar, but exhibit more communication at later times due to the additional horizontal fracture. A horizontal cross-section at $z=20$ m (Figures~\ref{fig::fracsC1C3FD}g-h) shows that the low level of numerical dispersion in 3D is comparable to that in 2D despite the large time-steps. 

For comparison, Figure~\ref{fig::fracsC1C3FD} shows IMPEC and IMPIC results for lower-order MHFE-FV simulations on Grids 1, 2, and 3. 
We note that on the same grid, the higher-order DG update takes less than twice the computational time as compared to FV (Table~\ref{table::ex4-performance}), but the numerical accuracy of DG is similar to that of FV on a grid with four times the number of elements.

\subsection{Example 4: CO$_{2}$ Injection into Multicomponent Oil on {Irregular} 2D and 3D Fractured Grids}
We extend the previous examples to multicomponent fluids. \co is injected into the same 5-component light oil as in the first example. Injection is from the bottom-left corner at a constant rate of 5\% PV/yr and production is from the diagonally opposite corner at constant pressure. We use the {irregular} 2D and 3D Grids 3 and 4 from the previous examples, with the same fracture and rock properties. We also consider a case in which the matrix permeability is 1 d, such that the contrast between fractures and matrix is reduced, and gravitational segregation in the vertical direction plays a more important role. 

Figure~\ref{fig::Compositional} summarizes the results for the \co molar fractions throughout the domain at 5\% and 40\% PVI. All simulations are carried out with the MHFE-DG, IMPIC approach. 
 First, we find that the composition profiles are somewhat similar to the methane injection case, but the \co front moves faster due to the higher viscosity ratio (Table~\ref{table::conditions}). We compare IMPIC simulations in which the time-step is chosen first as $100\times$ and then $5000\times$ the CFL condition, and find very similar low levels of numerical dispersion, suggesting that the time-step error is less than the spatial discretization errors. 
 {This is partly because in the \textit{matrix} the flow is slow, with a much larger local CFL condition than in the fractures.}
 As before, the 3D results show more communication due to the additional horizontal fracture, which intersects 6 of the vertical ones. Finally, when the matrix permeability is high (1 d), the effect of the fractures on the \co propagation front is significantly reduced and we see gravitational segregation of dense \co to the bottom. 

In terms of computational efficiency, we find that the 3D simulation for 6 components required only 16\% more CPU time than for the two-component C$_{1}$-C$_{3}$ case (Table~\ref{table::ex4-performance}). This is because only one large matrix needs to be inverted (Section~\ref{sec::numerical}), which is the same for each species. Each additional species only involves a much cheaper matrix multiplication. 

\subsection{Example 5: Viscous and Gravitational Fingering}
In this last example, we consider one of the scenarios in which numerical dispersion from lowest-order methods or large implicit time-steps can obscure important flow patterns. Specifically, we model gravitational instabilities (or fingering) that can occur when denser \co is injected from the top of a reservoir saturated with a lighter oil (20\% in Table~\ref{table::conditions}). \co also has an adverse viscosity ratio with respect to the oil ($\mu_\mathrm{oil}= 3.4\times \mu_{\mathrm{CO}_{2}}$), so flow is susceptible to viscous instabilities as well. We have investigated both instabilities, but will only present the modeling of gravitational fingering.

We consider $20\times 100\ \mathrm{m}^{2}$ and $20\times 5\times 100\ \mathrm{m}^{3}$ 2D and 3D (Grid 5) domains on $40\times (10\times)\ 200$ grids with a permeability of 10 md. \co is injected uniformly from the top at a rate of 5\% PV/yr, which is slow enough for density- or gravity-driven flow to be equal or larger than the viscous flow component. The density-driven flow is proportional to the density contrast between injected \co and oil in place \citep{xu06,riaz,philipsequestration}, and amplified by the adverse viscosity ratio (Table~\ref{table::conditions}).
 It is also proportional to the rock permeability: in highly permeable media the flow will be extremely unstable, whereas in tighter rock the growth rate and propagation speed of fingers is slow enough to be insignificant compared to pressure driven advective flow. The case considered here (10 md) is intermediate. 

In the absence of hydrodynamic dispersion, there is no \textit{physical} restoring force and the on-set and initial wavelength of the fingering instability is largely determined by \textit{numerical} dispersion. Without diffusion, or for typically small Fickian diffusion coefficients of order $10^{-8}\ \mathrm{m}^{2}/\mathrm{s}$, the gravitational instability should develop nearly instantly and with many small-scale fingers. Higher-order methods are able to resolve such features on coarse grids, while lowest-order methods require much finer grids, particularly in 3D. 

\begin{figure}[h!]
 \centering
\includegraphics[width=\textwidth]{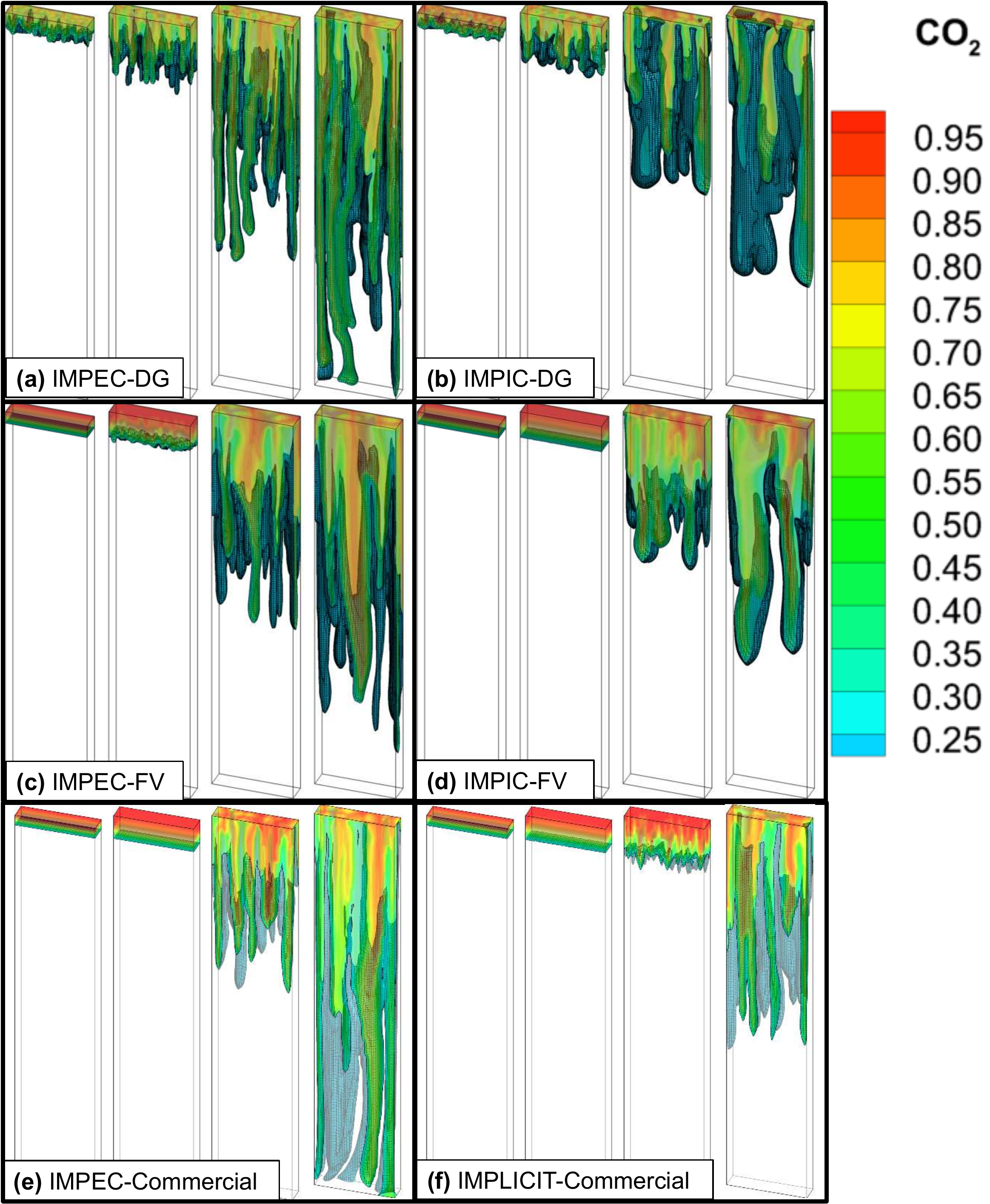} 
\caption{Example 5: CO$_{2}$ molar fraction at 2\%, 5\%, 15\% and 22\% PVI for four different (combinations of) numerical methods: IMPEC with a DG transport update ($\mathbf{a}$), IMPIC-DG with time-step sizes $100\times$ the CFL condition ($\mathbf{b}$); IMPEC with a FV transport updated ($\mathbf{c}$), and IMPIC-FV with time-step sizes $100\times$ the CFL condition ($\mathbf{d}$). Simulations with a commercial simulator using the IMPES ($\mathbf{e}$) and implicit ($\mathbf{f}$) options are presented as well.
\co is injected uniformly from the top at 5\% PV/yr, and production is at constant pressure from the bottom. The grid (Grid 5) is $20\times 5\times 100\ \mathrm{m}^{3}$ with $40\times 10\times 200$ elements.
}
 \label{fig::fingeringstruct}
 \end{figure}

Figure~\ref{fig::fingeringstruct} shows that DG simulations, using both an IMPEC scheme and the new IMPIC update with $100$ times larger time-steps can resolve the early small-scale on-set of the fingering instability at only 2\% PVI, while the front is still stable for both IMPEC and IMPIC lowest-order FV simulations (the latter with the same time-step sizes as IMPIC-DG). At later times, even FV simulations \textit{can} resolve the gravitational instability, which is quite pronounced due to the large density (and adverse viscosity) contrast. The FV simulations exhibit more numerical dispersion than DG due to the spatial discretization errors, while IMPIC simulations with much larger (backward Euler) time-steps result in an increasingly large ${\cal O}(\Delta t)$ temporal discretization error. While all simulations exhibit fingering behavior, finer fingers are resolved by IMPEC-DG, which results in early breakthrough at 20\% PVI. In the more disperse FV and IMPIC simulations, individual fingers may not be resolved, resulting in fewer and larger fingers and a later on-set. Most importantly, for practical applications, are the differences in breakthrough times: the more numerical dispersion, the later the breakthrough time of unstable fingers.

{Figure~\ref{fig::fingeringstruct} also includes simulation results from a widely used (lowest-order FV) commercial reservoir simulator (referred to as CS) with both its CS-IMPEC and implicit (CS-IMP) time-discretization options (with default tuning). We see that, while the results are qualitatively similar, the onset time of the fingering instability is delayed as compared to our FV results for both IMPEC and IMPIC. This is most likely due to our more accurate MHFE velocity field. At later times, the fingers appear to reach the bottom of the domain slightly earlier in the CS results. This is probably caused by differences in the phase behavior modeling, specifically the local CO$_{2}$-saturated oil density. In terms of CPU efficiency (Table~3), we find that our IMPEC-FV method has almost identical efficiency to the CS-IMPEC option. Our IMPIC-FV results are 32$\times$ faster than the CS-IMP simulator. This is partly due to our choice of time-step size, which is $3.5\times$ the average time-step size selected by the CS-IMP. More importantly, though, CS-IMP required $9.5\times$ more CPU time \textit{per time-step} than our implementation (though we acknowledge that this may be improved by tuning the CS-IMP iterative/convergence criteria). Our higher-order schemes further improve the efficiency.}

We perform a number of additional simulations of gravitational fingering with increasing complexity on 2D and 3D {irregular}, quadrilateral, triangular, hexahedral, and some fractured grids. The 3D grids are logically the same as in Figure~\ref{fig::fingeringstruct}, but with a sinusoidal variation in the $x$-direction (Grid 6), while the 2D quadrilateral grids (Grid 7) are the same as the 3D ones, but without the $y$-direction. The triangular grid (Grid 8) has $18,830$ elements.

We find that the on-set time and initial growth of the fingers in the 3D DG and FV simulations is qualitatively the same as for the structured 3D grids. At later times, as expected, the fingers reach the sloping right-side in the domain, where multiple fingers merge and flow down the incline towards the production well in the bottom. These results mainly demonstrate that our new numerical methods perform equally well on orthogonal and {irregular} grids, with the same computational efficiency and accuracy.

The 2D results show the highest resolution for IMPEC-DG simulations and significant numerical dispersion for both implicit (IMPIC) DG simulations when the time-step size is increased first 10- then 100-fold, as well as for IMPEC-FV on quadrilateral grids. The next two panels show robust performance of the MHFE-DG-CFE methods for highly non-linear flow instabilities, on {irregular} grids, and with discrete fractures (Grid 9). In these simulations the (matrix) permeability is not perturbed, but the on-set and growth of the instability is the same. Once multiple fingers reach the horizontal fracture, they flow through the fracture and `pour' out of the only outlet on the left-hand-side, forming a new gravitational finger. 

Finally, the last two panels show results for IMPEC and IMPIC DG simulations on a fine triangular grid. The exact fingering flow patterns are different from the quadrilateral grids, but the resolution is similar, both for IMPEC (compare to Figure~\ref{fig::fingeringstruct}c) and IMPIC (Figure~\ref{fig::fingeringstruct}e). A more detailed study of fingering behavior with our MHFE-DG methods is presented in \cite{moortgat2016viscous}.

The simulations in this example demonstrate the robustness and efficiency of our new higher-order IMPIC approach for complex multicomponent, compressible flow problems in discretely fractured reservoirs and on {irregular} 2D and 3D grids. 

\section{Conclusions}
We have developed lowest- and higher-order FE implicit-pressure-implicit-composition, or IMPIC, methods for unstructured triangular, quadrilateral (2D), and hexahedral (3D) grids. Discrete fractures are included through the cross-flow-equilibrium model, in which the fracture transport update is identical to that in the matrix. For fractured reservoirs, the unconditional stability of the IMPIC scheme allows for three orders of magnitude larger time-steps than the IMPEC approach while maintaining accuracy

\begin{figure}[h!]
 \hspace{-5em}
\includegraphics[width=1.3\textwidth]{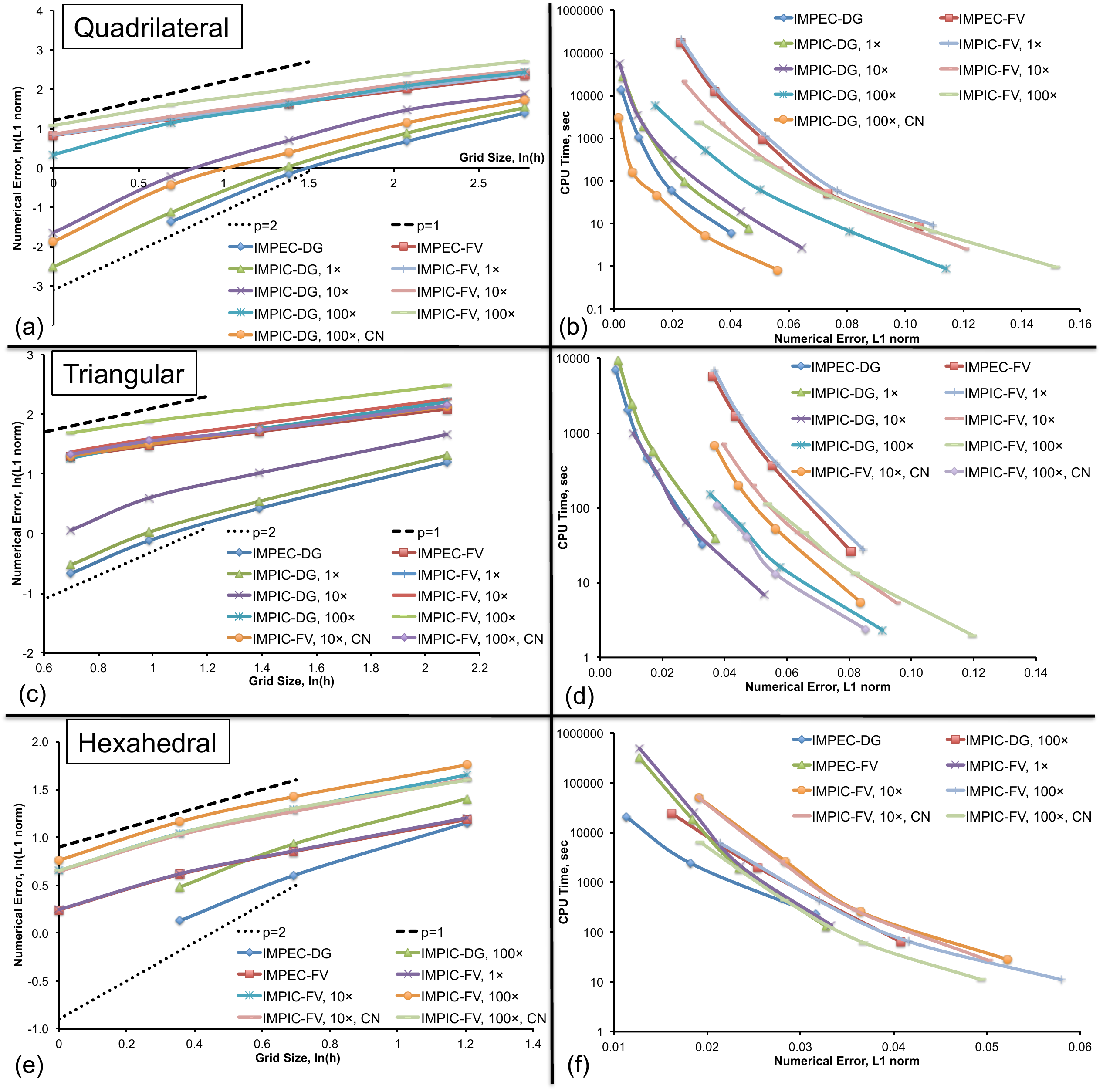} 
\caption{Example 1: $\mathrm{L}_{1}$ error versus grid size $h$ and CPU time versus $\mathrm{L}_{1}$ error on quadrilateral \textbf{(a)}--\textbf{(b)}, triangular \textbf{(c)}--\textbf{(d)}, and hexahedral \textbf{(e)}--\textbf{(f)} grids. CN denotes Crank-Nicolson, and IMPIC results are shown for $1\times$, $10\times$, and $100\times$ the CFL constraint. 
}
 \label{fig::ex1b}
 \end{figure}

A large number of test simulations were carried out to investigate the accuracy and efficiency for challenging problems. We  draw the following conclusions from the performance summaries in Figure~\ref{fig::ex1b} and Table~\ref{table::ex4-performance}: 
\begin{enumerate} 
\item The higher-order DG simulations using both IMPEC and IMPIC methods on 2D grids require remarkably little extra CPU time as compared to the lowest-order FV updates, while achieving much higher accuracy;
\item On 3D grids the matrices that need to be inverted in the IMPIC-DG update become very large ($0.4\times 10^{12}$ elements for Grids 5-6, most of which are zero) and the CPU time increases non-linearly, due to our use of a direct solver (Pardiso). Iterative solvers should provide higher computational efficiency for the same numerical methods;
\item For reasonable grid sizes, the implicit update is surprisingly efficient. As an example, for Grid 7 (quadrilateral), each IMPIC-DG update only requires 60\% more CPU time than a single IMPEC update, while 100 times larger time-steps are used, achieving a speed increase by a factor of $\sim 20$. For 3D FV simulations on hexahedral Grid 5 a speed-up of nearly $40\times$ is achieved by using $100\times$ larger time-steps, because each IMPIC update only requires $71\%$ more CPU time;
\item The aforementioned efficiency is partly due to the \textit{linear} direct solution method for the single-phase flow equations. For multiphase flow, even the decoupled equations become non-linear and iterative solution methods have to be used (e.g., Newton-Raphson). In such methods, multiple iterations have to be carried out per time-step, each of which requires at least as much CPU time as the direct solution in this work.  
\item While the implicit Euler methods presented in this work are highly efficient, in the sense of allowing larger time-steps than the IMPEC approach, the larger time-steps also result in increased numerical dispersion. To achieve a given accuracy, particularly on unfractured grids, the IMPEC approach may be more efficient than the IMPIC scheme (which would require much finer grids, or small time-steps);
\item Second-order Crank-Nicolson implicit time discretizations were developed as well, and significantly improve the performance of the IMPIC-DG method. For IMPIC-FV spatial errors may exceed temporal ones and the improvement from Crank-Nicolson is less pronounced; 
\item For fractured reservoirs, an implicit transport update, at least in the fractures, is almost unavoidable, especially in 3D. We demonstrated in Example~\ref{sec::ex2} that we can achieve both high accuracy and computational efficiency with our new higher-order DG implicit methods, combined with the cross-flow-equilibrium discrete fracture model. 
\end{enumerate}

These models are developed for multicomponent single-phase compressible flow and provide a highly efficient scheme to study first-contact-miscible gas injection into fractured gas and oil reservoirs, as well as multicomponent contaminant transport in an aqueous phase, allowing for density driven flow.

 \begin{table*}[htdp]
 \caption{\label{table::FCMcomp}Composition (molar fraction) $z_i^0$, acentric factor $\omega$, critical temperature $T_c$, critical pressure $p_c$, molar weight $M_w$, critical volume $V_c $ and volume translation $s$ in fluid characterization.}
 \begin{center}
 \begin{tabular*}{\hsize}{@{\extracolsep{\fill}}lrrrrrrr}\hline
 Species &  $z_i^0$ & $\omega$ &$T_c (\mathrm{K})$ & $p_c (\mathrm{bar})$  & $M_w (\mathrm{g}/\mathrm{mole})$& $V_c \left(\mathrm{cm}^3/\mathrm{g}\right)$ & $s$\\\hline
$\mathrm{CO}_2$						& 0.000	&   0.239	&         304	&          74	&          44	&    2.14	&  -0.177\\
$\mathrm{C}_1\! +\! \mathrm{N}_2$		& 0.567	&   0.012	&         189	&          46	&          16	&    6.09	&  -0.157\\
$\mathrm{C}_2\! -\! \mathrm{C}_3$		& 0.155	&   0.120	&         330	&          46	&          35	&    4.73	&  -0.094\\
$\mathrm{C}_4\! -\! \mathrm{C}_6$		& 0.079	&   0.233	&         455	&          35	&          69	&    4.32	&  -0.048\\
$\mathrm{C}_7\! -\! \mathrm{C}_{10}$		& 0.091	&   0.428	&         584	&          24	&         120	&    4.25	&   0.055\\
$\mathrm{C}_{11+}$						& 0.108	&   1.062	&         751	&          13	&         293	&    4.10	&   0.130\\
 \hline
 \end{tabular*}
 \end{center}
 \end{table*}
 
  \begin{table*}[htdp]
 \caption{\label{table::conditions}Fluid properties at given reservoir temperature $T$ and pressure $p$. Specifically the mass density and viscosity of the initial reservoir fluid, the injected fluid, and a 50:50 mol\% mixture of the two fluids.}
 \begin{center}
 \begin{tabular*}{\hsize}{@{\extracolsep{\fill}}llrrr||ll}\hline
Ex. & Property & Initial & Injected & Mixture &  Conditions \\\hline
2--3 & Density ($\mathrm{kg}/\mathrm{m}^{3}$) & 120 & 25 & 53 & $T = 124^{\circ}$ C \\ 
2--3 & Viscosity (cp) & 0.015 & 0.015 & 0.014 & $p = 50$ bar \\\hline 
1 and 4--5 & Density ($\mathrm{kg}/\mathrm{m}^{3}$) & 571 & 683 & 612 & $T = 127^{\circ}$ C \\ 
1 and 4--5& Viscosity (cp) & 0.175 & 0.051& 0.114 & $p = 483$ bar \\\hline   \end{tabular*}
 \end{center}
 \end{table*}

  \begin{table*}[htdp]
 \caption{\label{table::ex4-performance}Examples 2-5: Performance Analyses using different methods and Grids: total CPU time, number of time-steps ($\Delta t$), CPU time per time-step. CS-IMPEC and CS-IMP refer to the IMPEC and implicit options of a commercial reservoir simulator, respectively.}
 \begin{center}{\footnotesize
 \begin{tabular*}{\hsize}{@{\extracolsep{\fill}}|lllrrrr|}\hline
 Ex. & Method & Grid & ($\times$CFL) & CPU time (s) & nr.~of $\Delta t$ & CPU/$\Delta t$ (s) \\\hline
2 & DG-IMPEC &  1 & 1& 3,454 & 268,925 & $12.8\times 10^{-3}$ \\
2 & DG-IMPIC &  1 & 1000& 8 & 318 & $25.2\times 10^{-3}$ \\
2 & DG-IMPEC &  2 & 1& 45,947 & 271,433 & 0.17 \\
2 & DG-IMPIC &  2 & 5000 & 55 & 103 & 0.53 \\\hline
3 & DG-IMPIC &  3 &  5000 & 48 & 105 & 0.46 \\
3 & DG-IMPIC &  4 & 5000 & 11,461 & 118 & 97.13 \\
3 & FV-IMPEC &  1 & 1 & 2,864 &  254,863 & $12.2\times 10^{-3}$ \\
3 & FV-IMPIC &  1 & 1000 & 5 & 304 & $16.4\times 10^{-3}$ \\
3 & FV-IMPIC &  2 &5000 & 32 & 100 & 0.32 \\
3 & FV-IMPIC &  3 & 5000 & 47 & 102 & 0.46 \\\hline
4 & DG-IMPIC &  3 & 100 & 2,558 & 3,217 & 0.80 \\
4 & DG-IMPIC &  3 & 5000 & 92& 110 & 0.84 \\
4 & DG-IMPIC &  4 & 5000 & 13,314& 128 & 104.0 \\
4 & DG-IMPIC &  4 & 5000 & 7,302 & 59 & 123.8 \\\hline
5 & DG-IMPEC &  5 & 1 & 31,134 & 8,557 & 3.64 \\
5 & DG-IMPEC &  5 & 100 & 24,621 & 131 & 187.9 \\
5 & FV-IMPEC &  5 & 1 & 26,828 & 8,501 & 3.16 \\
5 & FV-IMPIC &  5 & 100 & 708 & 131 & 5.40 \\
5 & CS-IMPEC & 5 &  - & 27,231 & 25,574 &   1.06 \\
5 & CS-IMP	& 5  & - & 23,343 & 453 & 51.53 \\
5 & DG-IMPEC &  6 & 1 & 31,706 & 8,749  & 3.62 \\
5 & FV-IMPEC &  6 & 1 & 26,873 & 8,510  & 3.16 \\
5 & DG-IMPEC &  7 & 1 	 & 403 & 1,848  & 0.22 \\
5 & DG-IMPIC &  7 & 10 	 & 71 & 227  & 0.31 \\
5 & DG-IMPIC &  7 & 100 & 19 & 55  & 0.35 \\
5 & FV-IMPEC &  7 & 1 & 307 & 1,892  & 0.16 \\
5 & DG-IMPEC &  8 & 1 & 919 & 3,488   & 0.26 \\
5 & DG-IMPEC &  9 & 1 & 9,906 & 20,597   & 0.48 \\
5 & DG-IMPIC &  9 & 100 & 150 & 252   & 0.60 \\
 \hline
 \end{tabular*}}
 \end{center}
 \end{table*}

\begin{figure}[h!]
 \centering
\includegraphics[width=\textwidth]{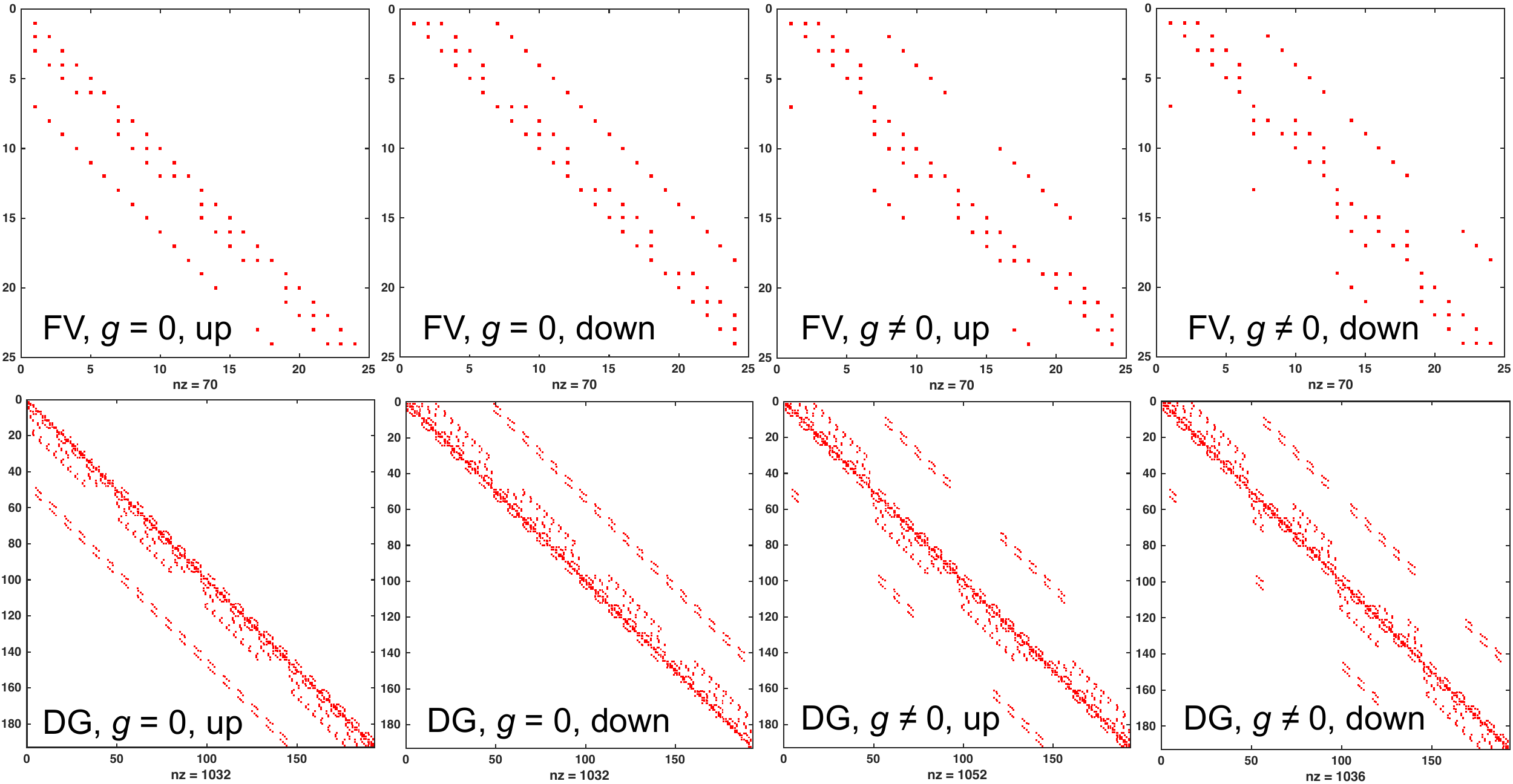} 
\caption{{Sparsity patterns for IMPIC-FV and IMPIC-DG on $2\ \times\ 3\ \times 4$ element hexahedral grids with injection either from element 1 or element 24 and production from the opposite corner, and both without and with gravity. The number of non-zero components is $nz = 70$ for IMPIC-FV and varies from 1,032 to 1,052 for the IMPIC-DG simulations.}}
 \label{fig::dofquad}
 \end{figure}

   \begin{figure}[h!]
 \centering
\includegraphics[width=\textwidth]{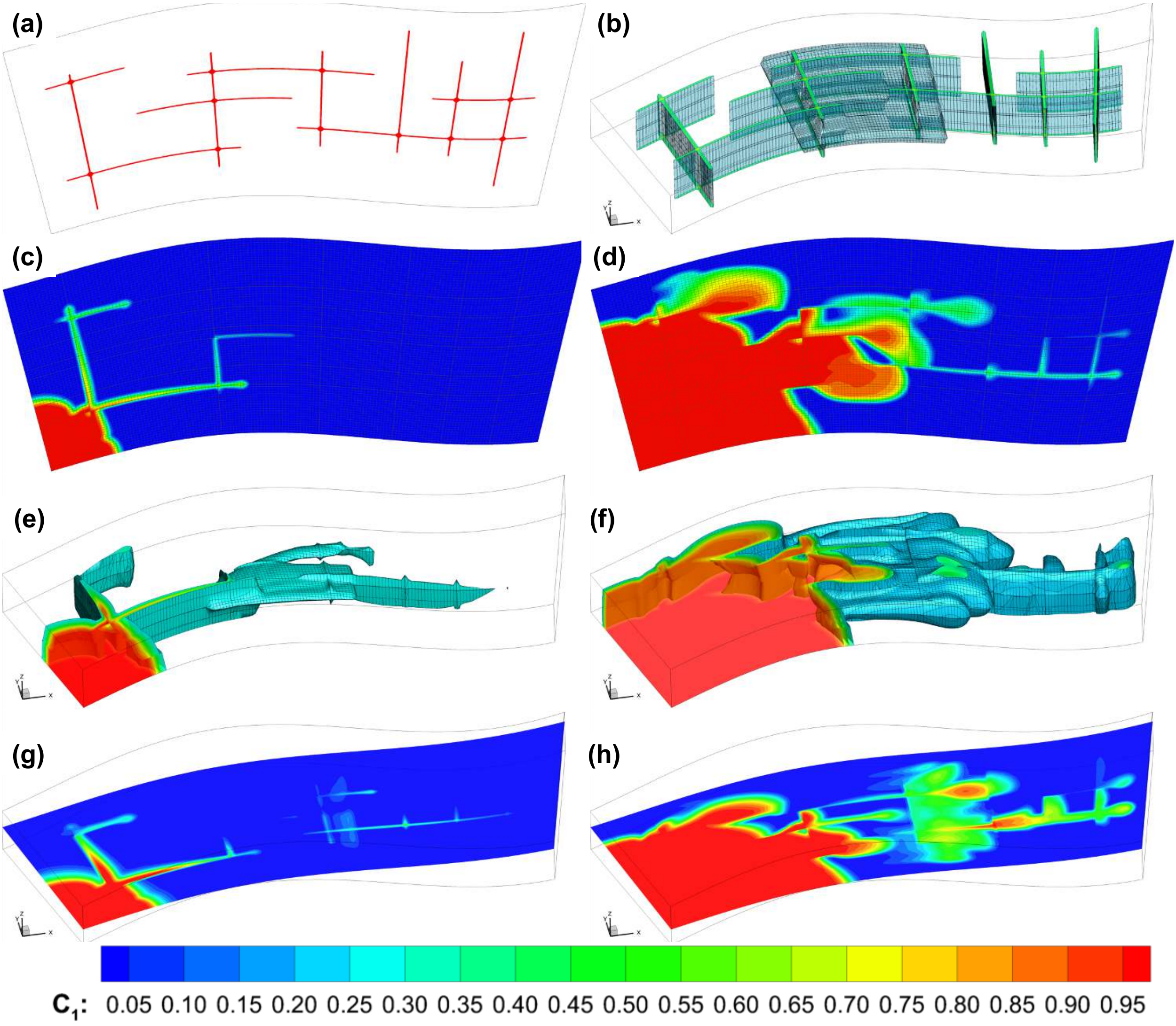} 
\caption{Example 3: Methane molar fraction at 5\% (left column) and 40\% (right column) PVI. The simulation set-up is as in Figure~\ref{fig::fracsC1C3}, but for {irregular} $166\times 86$ 2D Grid 3 ($\mathbf{a}$) and $166\times 86\times 10$ 3D Grid 4 ($\mathbf{b}$). The 3D grid has an additional horizontal fracture in addition to the 12 discrete vertical fractures in the 2D grid. Methane molar fraction is shown for the 2D grid (at 5\% ($\mathbf{c}$) and 40\% ($\mathbf{d}$) PVI), for the 3D grid (at 5\% ($\mathbf{e}$) and 40\% ($\mathbf{f}$) PVI) and for a horizontal cross-section through the 3D grid at $z=20$ m (at 5\% ($\mathbf{g}$) and 40\% ($\mathbf{h}$) PVI). Results are for MHFE-DG implicit simulations with $5000\times $CFL.}
 \label{fig::fracsC1C3unstr}
 \end{figure}
 
    \begin{figure}[h!]
 \centering
\includegraphics[width=\textwidth]{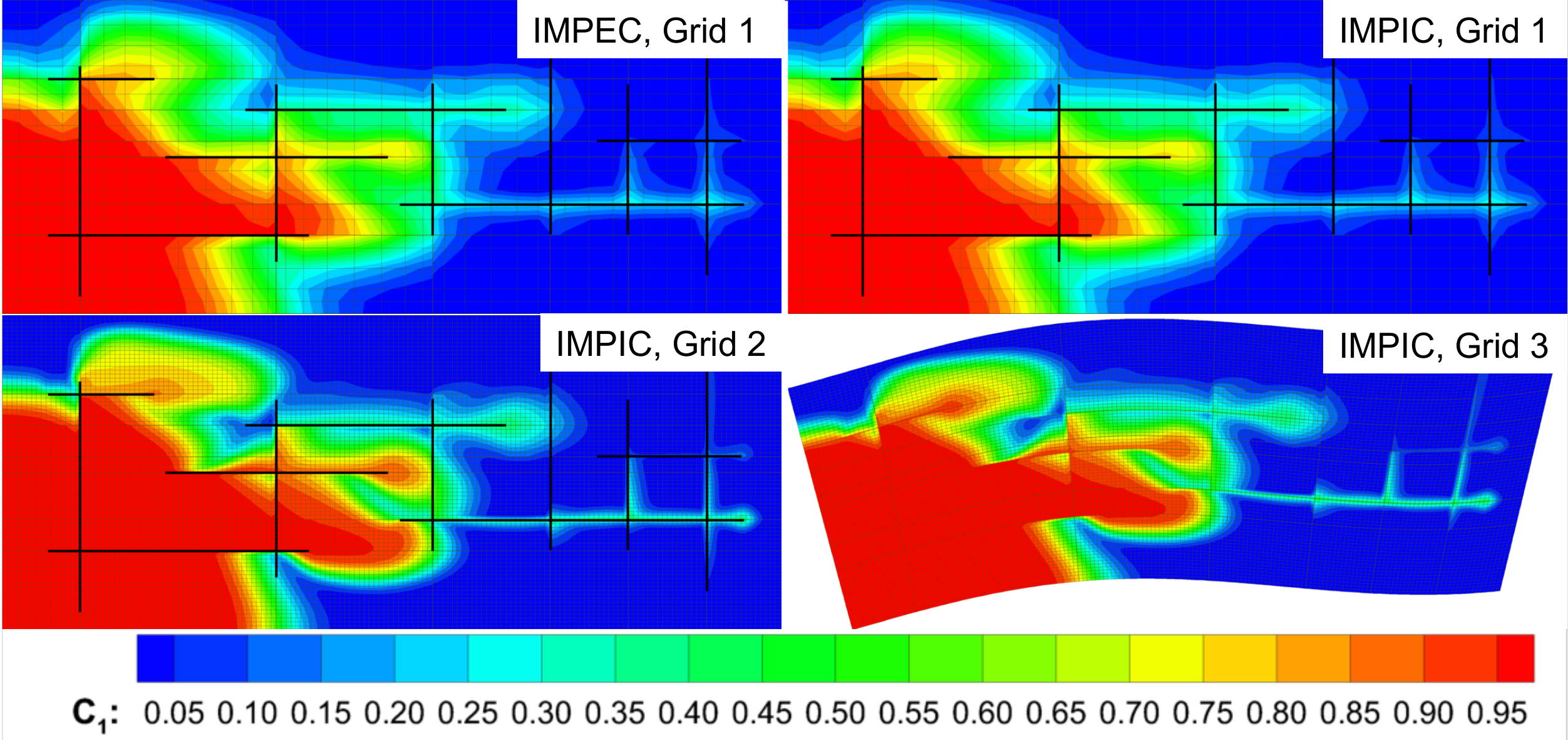} 
\caption{Example 3: Methane molar fraction at 40\% PVI. The simulation set-up is as in Figures~\ref{fig::fracsC1C3} and \ref{fig::fracsC1C3unstr}, but for lower-order MHFE-FV simulations. Simulations are on the coarse Grid 1, comparing IMPEC and IMPIC, as well as IMPIC simulations on the finer Grid 2, and the {irregular} Grid 3. All implicit time-steps are $1000\times$CFL.}
 \label{fig::fracsC1C3FD}
 \end{figure}

    \begin{figure}[h!]
 \centering
\includegraphics[width=\textwidth]{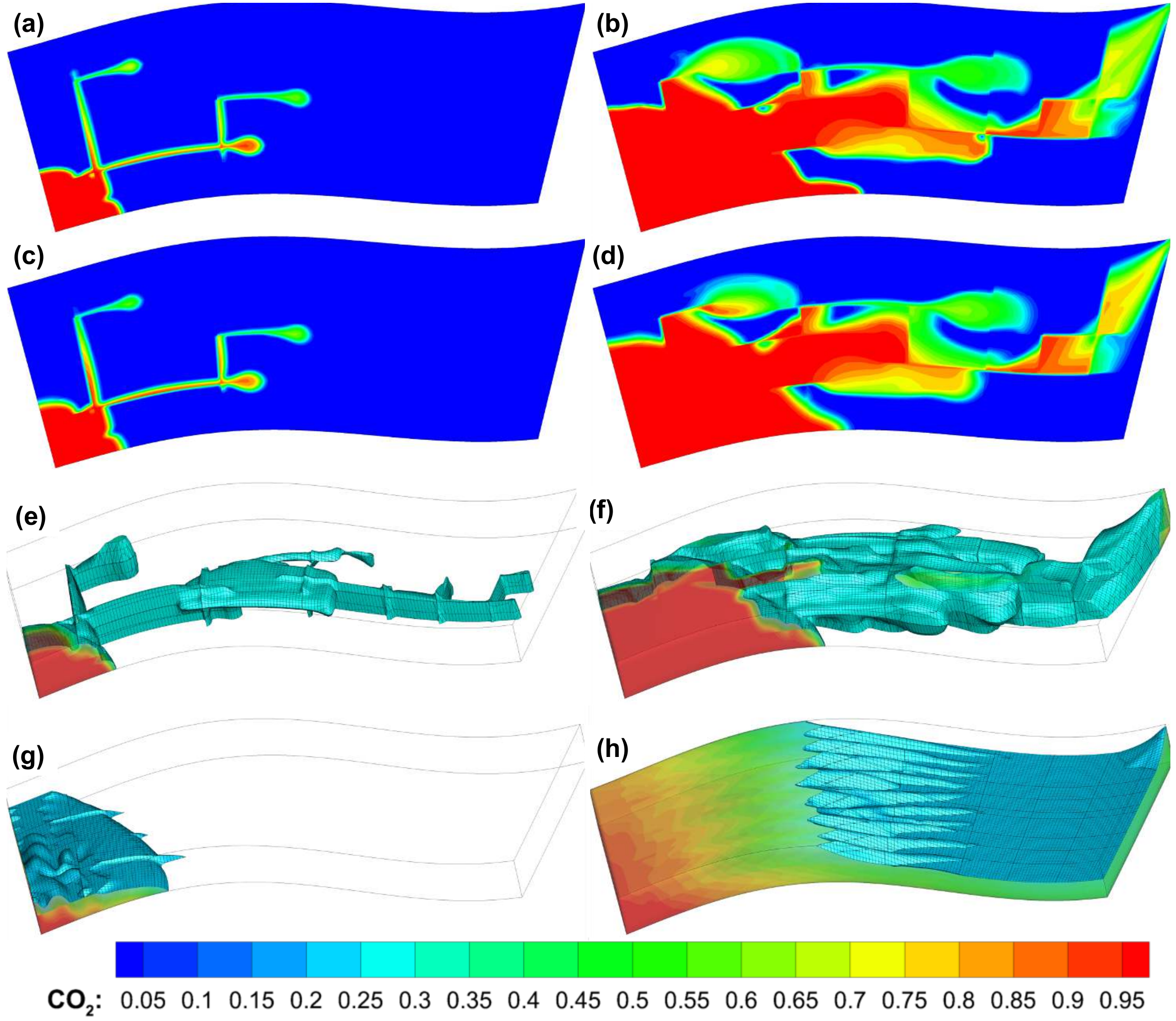} 
\caption{Example 4: CO$_{2}$ molar fraction at 5\% (left column) and 40\% (right column) PVI. Grid and fracture locations are as in Figure~\ref{fig::fracsC1C3unstr}, but in these simulations CO$_{2}$ is injected (at the same rate and well locations) into a reservoir saturated with a 5 (pseudo)-component oil. Results are for MHFE-DG implicit simulations, performed in 2D with both $100\times$ ($\mathbf{a}$ and $\mathbf{b}$) and $5000\times$ ($\mathbf{c}$ and $\mathbf{d}$) the CFL time-step size, and in 3D ($\mathbf{e}$ and $\mathbf{f}$), also with a factor $5000\times $CFL. The 3D simulations are repeated with a $1000\times$ higher matrix permeability of 1d to show a more pronounced effect of gravity ($\mathbf{g}$ and $\mathbf{h}$).}
 \label{fig::Compositional}
 \end{figure}
 
      \begin{figure}[h!]
 \centering
\includegraphics[width=\textwidth]{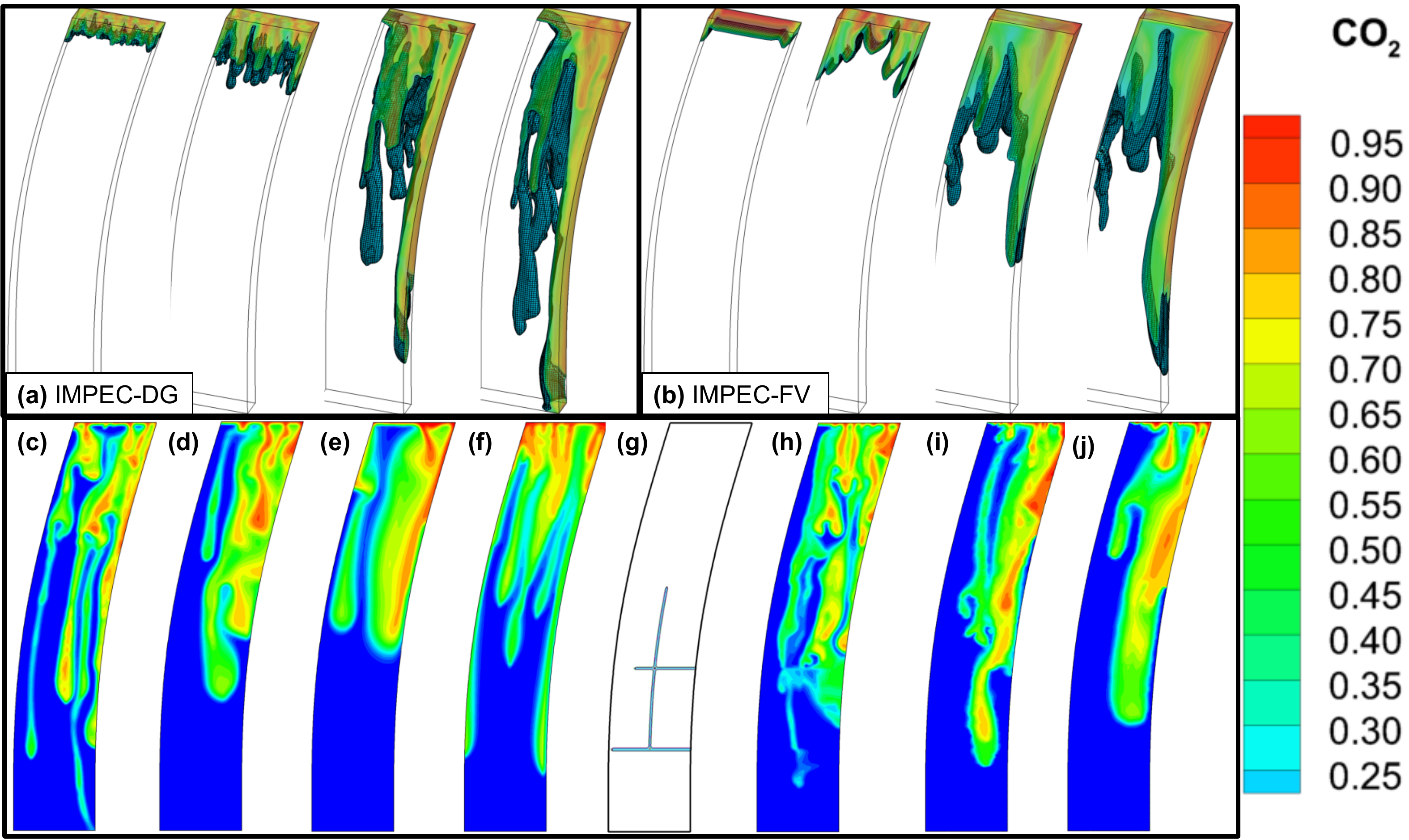} 
\caption{Example 5: CO$_{2}$ molar fraction at 2\%, 5\%, 15\% and 22\% PVI for IMPEC DG ($\mathbf{a}$) and FV ($\mathbf{b}$) 3D (Grid 6) simulations with the same set-up as in Figure~\ref{fig::fingeringstruct}. Lower panels show CO$_{2}$ molar fraction at 20\% PVI 2D (Grid 7) simulations on a $40\times 200$ quadrilateral grid using different numerical methods: IMPEC-DG ($\mathbf{c}$), IMPIC-DG with $10\times$ CFL time-steps ($\mathbf{d}$), IMPIC-DG with $100\times$ CFL time-steps ($\mathbf{e}$), IMPEC-FV ($\mathbf{f}$), with 2 discrete fractures ($\mathbf{g}$) using IMPEC-DG, but without a matrix permeability perturbation ($\mathbf{h}$). Two more simulation results are shown for a $18,830$ element triangular grid with IMPEC ($\mathbf{i}$), and IMPIC with $100\times$ CFL time-steps ($\mathbf{j}$). 
}
 \label{fig::fingeringunstruct}
 \end{figure}
 
 \newpage
 
 \appendix
\section{Matrix Form of Tri-Linear Discontinuous Galerkin Transport Update on Hexahedral Grid}
For facilitate the implementation of the implicit DG transport update, we provide the explicit algebraic expressions in Eq.~(\ref{eq::dgupdate}) for the most complicated, hexahedral, elements. The numbering of nodes and faces, implied in the following equations, is illustrated in \cite{moortgatIX}. Nodes are numbered counter-clock-wise from the bottom-left-front corner (and again from the top-left-front corner for the upper face of a hexahedron).
 
 We drop the species index $i$ for clarity (the equations are the same for each species), as well as the element index $K$, and define vectors $\mathbf{q}$ of all $6$ face fluxes $q_{E}$, and $\mathbf{c}$ of the $8$ nodal molar densities (\textit{inside} element $K$). With this notation, the discretized volume integral in Eq.~(\ref{eq::dgupdate}) can be written as
 \begin{equation}
 F(c_{N},q_{E}) = (\mathbf{V}_{N} \mathbf{q}) \mathbf{c},
 \end{equation} 
 where each node is updated with a different matrix $\mathbf{V}_{N}$, as defined below (zeroes are indicated by `$\ast$' for clearer visualization of the non-zero elements).
 
 {\footnotesize
  \hspace{-5em}
\begin{eqnarray}\nonumber
\bar{\vec{V}}_{1} =
 \left( \begin{array}{cccccc}  
 -1 & 2 & -1 & 2 & -1 & 2 \\
 -2 & 1 & \ast & \ast & \ast & \ast \\
 \ast & \ast & \ast & \ast & \ast & \ast \\
 \ast & \ast & -2 & 1 & \ast & \ast \\
  \ast & \ast & \ast & \ast & -2 & 1 \\
 \ast & \ast & \ast & \ast & \ast & \ast \\
 \ast & \ast & \ast & \ast & \ast & \ast \\
 \ast & \ast & \ast & \ast & \ast & \ast 
   \end{array} \right), 
   \bar{\vec{V}}_{2} =
 \left( \begin{array}{cccccc}  
1 & -2 & \ast & \ast & \ast & \ast \\
2 & -1 & -1 & 2 & -1 & 2\\
\ast & \ast & -2 & 1 & \ast &  \ast\\
 \ast & \ast & \ast & \ast & \ast & \ast \\
 \ast & \ast & \ast & \ast & \ast & \ast \\
  \ast & \ast & \ast & \ast & -2 & 1\\
 \ast & \ast & \ast & \ast & \ast & \ast \\
 \ast & \ast & \ast & \ast & \ast & \ast 
    \end{array} \right),\bar{\vec{V}}_{3} =
 \left( \begin{array}{cccccc}  
  \ast & \ast & \ast & \ast & \ast & \ast \\
\ast & \ast & 1 & -2 & \ast & \ast \\
2 & -1 & 2 & -1 & -1 & 2 \\
1 & -2 & \ast & \ast & \ast & \ast \\
 \ast & \ast & \ast & \ast & \ast & \ast \\
 \ast & \ast & \ast & \ast & \ast & \ast \\
\ast & \ast & \ast & \ast & -2 & 1\\
 \ast & \ast & \ast & \ast & \ast & \ast 
   \end{array} \right), \\\nonumber
    \end{eqnarray}}
    
  {\footnotesize
\hspace{-5em}
 \begin{eqnarray}\nonumber
   \bar{\vec{V}}_{4} =
 \left( \begin{array}{cccccc}  
 \ast & \ast & 1 & -2 & \ast & \ast\\
  \ast & \ast & \ast & \ast & \ast & \ast \\
-2 & 1 & \ast & \ast & \ast & \ast \\
-1 & 2 & 2 & - 1 & -1 & 2 \\
  \ast & \ast & \ast & \ast & \ast & \ast \\
  \ast & \ast & \ast & \ast & \ast & \ast \\
  \ast & \ast & \ast & \ast & \ast & \ast \\
 \ast & \ast & \ast & \ast & -2 & 1 
    \end{array} \right), 
\bar{\vec{V}}_{5} =
 \left( \begin{array}{cccccc}  
   \ast & \ast & \ast & \ast &1 & -2\\
    \ast & \ast & \ast & \ast & \ast & \ast \\
  \ast & \ast & \ast & \ast & \ast & \ast \\
  \ast & \ast & \ast & \ast & \ast & \ast \\
-1 & 2 &-1 & 2 & 2 & -1\\
 -2 & 1 & \ast & \ast & \ast & \ast \\
   \ast & \ast & \ast & \ast & \ast & \ast \\
\ast & \ast & -2 & 1 & \ast & \ast
   \end{array} \right),  
   \bar{\vec{V}}_{6} =
 \left( \begin{array}{cccccc} 
   \ast & \ast & \ast & \ast & \ast & \ast \\
   \ast & \ast & \ast & \ast & 1 & -2\\
    \ast & \ast & \ast & \ast & \ast & \ast \\
  \ast & \ast & \ast & \ast & \ast & \ast \\
  1 & -2  & \ast & \ast & \ast & \ast\\
 2 & -1 & -1 & 2 & 2 & -1 \\
 \ast & \ast & -2 & 1 & \ast & \ast\\
    \ast & \ast & \ast & \ast & \ast & \ast 
    \end{array} \right), \\\nonumber
    \end{eqnarray}
    
 \begin{eqnarray}\nonumber
\bar{\vec{V}}_{7} =
 \left( \begin{array}{cccccc}  
    \ast & \ast & \ast & \ast & \ast & \ast \\
  \ast & \ast & \ast & \ast & \ast & \ast \\
   \ast & \ast & \ast & \ast & 1 & -2 \\
      \ast & \ast & \ast & \ast & \ast & \ast \\
  \ast & \ast & \ast & \ast & \ast & \ast \\
  \ast & \ast & 1 & -2 & \ast & \ast \\
  2 & -1 & 2 & -1 & 2 & -1 \\
  1 & -2 & \ast & \ast & \ast & \ast
   \end{array} \right), \quad 
   \bar{\vec{V}}_{8} =
 \left( \begin{array}{cccccc} 
   \ast & \ast & \ast & \ast & \ast & \ast \\
      \ast & \ast & \ast & \ast & \ast & \ast \\
   \ast & \ast & \ast & \ast & \ast & \ast \\
   \ast & \ast & \ast & \ast &1 & -2 \\
   \ast & \ast & 1 & -2 & \ast & \ast \\
      \ast & \ast & \ast & \ast &   \ast & \ast \\
   -2 & 1 & \ast & \ast &   \ast & \ast \\
-1 & 2 & 2 & -1 & 2 & -1
    \end{array} \right).
    \end{eqnarray}}
 
The DG discretized surface integral $G(c_{N},q_{E})$ in Eq.~(\ref{eq::dgupdate}) is the most complicated: the update of the molar density at each node $N$ involves the fluxes through all 6 faces, but only one nodal density on each face. Moreover, due to the discontinuous nature of the DG method, fluid properties have different values on either side of each face. Each \textit{global} node is shared by 8 grid cells (for a hexahedral grid), each of which can have a different \textit{local} value (of molar densities) \textit{inside} each grid cell sharing that global node. 

This complicates the upwinding procedure. Consider node 1 in a hexahedron: the upwind value with respect to a flux through edge 2 (left) is given by node 2 in the left-neighboring element. The upwind value at that same node 1, though, with respect to flux 6 (bottom face), will come from node 5 of the bottom (in $z$) neighbor. And, finally, the upwind value with respect to flux 4 (front face) is from node 4 from the front-neighboring element in the $y$-direction. To construct a global matrix that automatically does the upwinding, we therefore define two `node-map' matrices: $\mathbf{N}$ and $\mathbf{N^{\prime}}$. When updating $c_{N}$, row $N$ of matrix $\mathbf{N}$ provides the local node numbers inside element $K$ that are multiplied with each flux $q_{E}\ge 0 $, while row $N$ of matrix $\mathbf{N}^{\prime}$ provides the equivalent local node numbers inside element $K^{\prime}$, neighboring face $E$, when $q_{E}<0$. This choice of upwind $c_{N,E}$ with respect to $q_{E}$ is denoted by $\widetilde{c_{\tilde{N},E}}$, while the weight-factor of each node with respect to face $E$ is given by the $8\times 6$ matrix $\mathbf{e}$.
With these definitions, the upwind contribution to the transport update can be constructed as:

 \begin{equation}
 G(c_{N},q_{E}) = 2 \sum_{E} e_{N,E} \widetilde{c_{\tilde{N},E}} q_{E}
 \end{equation} 
 with 
 {\footnotesize
 \begin{eqnarray}\nonumber
 \mathrm{e} = 
 \left(
 \begin{array}{cccccc}
 -1 & 2 & -1 & 2 & -1 & 2 \\
 2  &  -1 &  -1 &  2 &  -1 &  2\\
 2  &  -1 &  2 &  -1 &  -1 &  2\\
 -1 &  2 &  2 &  -1 &  -1 &  2\\
 -1 &  2 &  -1 &  1 &  2 &  -1\\
 2 &  -1 &  -1 &  2 &  2 &  -1\\
 2 &  -1 &  2 &  -1 &  2 &  -1\\
-1 &  2 &  2 &  -1 &  2 &  -1
 \end{array}\right),  \mathbf{N} = 
 \left(
 \begin{array}{cccccc}
 2  &  1  &  4  &  1  &  5  &  1 \\
 2  &  1  &  3  &  2  &  6  &  2 \\
3  &  4  &  3  &  2  &  7  &  3 \\
3  &  4  &  4  &  1  &  8  &  4 \\
6  &  5  &  8  &  5  &  5  &  1 \\
6  &  5  &  7  &  6  &  6  &  2 \\
7  &  8  &  7  &  6  &  7  &  3 \\
7  &  8  &  8  &  5  &  8  &  4 
 \end{array}\right), 
   \mathbf{N^{\prime}} = 
 \left(
 \begin{array}{cccccc}  
1  &  2  &  1  &  4  &  1  &  5 \\
1  &  2  &  2  &  3  &  2  &  6 \\
4  &  3  &  2  &  3  &  3  &  7 \\
4  &  3  &  1  &  4  &  4  &  8 \\
5  &  6  &  5  &  8  &  1  &  5 \\
5  &  6  &  6  &  7  &  2  &  6 \\
8  &  7  &  6  &  7  &  3  &  7 \\
8  &  7  &  5  &  8  &  4  &  8 
 \end{array}\right) 
 \end{eqnarray}}
 
 The expressions for triangular, quadrilateral, and other types of grid elements proceed along the same lines.

{Figure~\ref{fig::dofquad} illustrates the sparsity pattern for both IMPIC-FV and IMPIC-DG simulations on hexahedral grids. The IMPIC-FV simulations are for a $2\ \times\ 3\ \times 4\ \mathrm{m}^{3}$ grid, with grid sizes of 1 m in each direction. Methane is injected into propane either from the grid origin (denoted as `up') or from the diagonally opposite corner (denote as `down'), which production in each case from the opposite corner. The sparsity patterns are shown at $50\%$ PVI for both simulations without and with gravity. Both gravity and general well locations result in non-symmetric matrices.}

\end{document}